\documentclass[aps,prc,reprint, longbibliography,notitlepage,nofootinbib,nobibnotes,superscriptaddress,amsmath,amssymb,preprintnumbers]{revtex4-2}

\usepackage{graphicx}
\usepackage{dcolumn}
\usepackage{bm}       
\usepackage{amssymb}  
\usepackage{adjustbox} 
\usepackage{amsmath,array}
\usepackage{mathtools}
\usepackage{comment}
\usepackage{natbib}
\usepackage{MnSymbol}
\usepackage{caption}

\usepackage{subcaption}

\usepackage[colorlinks=true,linkcolor=blue,citecolor=blue, urlcolor=blue]{hyperref}
\usepackage[capitalise]{cleveref}

\usepackage[usenames,dvipsnames]{color}

\usepackage     [utf8]                  {inputenc}
\usepackage     [T1]                    {fontenc}
\usepackage     [english]               {babel}

\usepackage{epigraph}
\usepackage{etoolbox}
\usepackage{ dsfont }
\usepackage{physics}
\usepackage{float}
\usepackage{xcolor}
\usepackage{siunitx}
\makeatletter
\patchcmd{\epigraph}{\@epitext{#1}}{\itshape\@epitext{#1}}{}{}  
\newcommand*\eqsize{%
\@setfontsize\mysize{9.0}{9.0}%
    }
    \usepackage{multirow}
\makeatother

\usepackage{xfrac}
\usepackage{amsmath,amssymb}
\usepackage{esdiff} 
\usepackage{commath} 
\usepackage{bbm} 
\usepackage{braket}
\usepackage{slashed}
\usepackage{tikz}

\usetikzlibrary{plotmarks,fadings,positioning,decorations.pathmorphing,decorations.markings,patterns}
\usepackage{pgfplots}
\pgfplotsset{compat=1.18}

\definecolor{oscar}{RGB}{22, 156, 172}
\definecolor{YHCOLOR}{RGB}{255, 119, 8}
\definecolor{darkgreen}{RGB}{50, 160, 80}

\newcommand{\rmd}{\mathrm{d}}
\newcommand{\rmi}{\mathrm{i}}
\newcommand{\rme}{\mathrm{e}}
\newcommand{\bfDelta}{\mathbf{\Delta}}

\newcommand{\ob}[1]{\mkern 1.5mu\overline{\mkern-1.5mu#1\mkern-1.5mu}\mkern 1.5mu}

\newcommand{\bt}{\mathbf{b}}

\newcommand{\bbt}{\mathbf{\ob{b}}}

\newcommand{\kt}{\mathbf{k}}
\newcommand{\zt}{\mathbf{z}}
\newcommand{\xt}{\mathbf{x}}
\newcommand{\yt}{\mathbf{y}}
\newcommand{\rt}{\mathbf{r}}
\newcommand{\rbt}{\mathbf{\ob{r}}}

\newcommand{\xbt}{{\ob{\xt}}}
\newcommand{\ybt}{{\ob{\yt}}}
\newcommand{\kbt}{{\ob{\kt}}}
\newcommand{\zbt}{{\ob{\zt}}}

\newcommand{\nn}{\nonumber}

\newcommand{\qqbar}{$\mathrm{q}\bar{\mathrm{q}}$}
\newcommand{\up}{^\mathrm}
\newcommand{\down}{_\mathrm}

\newcommand{\Nc}{N\down{c}}
\newcommand{\CF}{C\down{F}}
\newcommand{\NH}{N\down{h}}

\DeclarePairedDelimiter\biabs{|}{|}
\DeclarePairedDelimiter\bigabs{\big |}{\big |}

\DeclarePairedDelimiter\avg{\langle}{\rangle}
\DeclarePairedDelimiter\bigavg{\big \langle}{\big \rangle}
\DeclarePairedDelimiter\biggavg{\Big \langle}{\Big \rangle}
\DeclarePairedDelimiter\bigggavg{\bigg \langle}{\bigg \rangle}
\DeclarePairedDelimiter\biggggavg{\Bigg \langle}{\Bigg \rangle}

\begin{document}

\date{\today}
\title{Saturation effects in exclusive vector meson production in DIS}
\author{Oscar Garcia-Montero}
\email{garcia@physik.uni-bielefeld.de}
\affiliation{Fakult\"at f\"ur Physik, Universit\"at Bielefeld, D-33615 Bielefeld, Germany}
\affiliation{Instituto Galego de F\'isica de Altas Enerx\'ias IGFAE, Universidade de Santiago de Compostela, E-15782 Galicia, Spain}

\author{Yannik Hoffmann}
\email{physics@yhoffmann.eu}
\affiliation{Fakult\"at f\"ur Physik, Universit\"at Bielefeld, D-33615 Bielefeld, Germany}

\author{Sören Schlichting}
\affiliation{Fakult\"at f\"ur Physik, Universit\"at Bielefeld, D-33615 Bielefeld, Germany}

\begin{abstract}
We investigate saturation effects in exclusive vector meson production in deep inelastic scattering (DIS), where we model fluctuations within the target protons as localized color-charge hotspots. Based on the Color Glass Condensate (CGC) framework and the dipole picture for vector meson production, we examine the dependencies of coherent and incoherent scattering cross sections on the momentum transfer. We draw conclusions on the effectiveness of our hot spot model and the strength of the suppression of the scattering cross sections caused by saturation effects. We find that saturation has mild effects in the given energy and charge-density ranges, but can also show that suppression becomes more prominent as the color-charge density inside the proton increases.

\end{abstract}

\maketitle
\section{Introduction} \label{sec:introduction}

Exclusive vector meson production in electron-proton ($\mathrm{e}\mathrm{p}$) scattering allows for unique insights into the internal structure of the proton and the underlying structure of Quantum Chromodynamics (QCD). Traditionally, research has focused on the proton's structure functions, which reveal properties of the parton distribution functions (PDFs)~\cite{Klasen:2023uqj}. Substantial experimental and theoretical work has been done in the past decades to extract such functions~\cite{Harland-Lang:2014zoa,NNPDF:2017mvq}.
Nevertheless, recent evidence gained from experiments on collective flow in proton-proton and proton-nucleus collisions indicates that the transverse profile of partons is also highly significant to describe observed spectra and particle correlations in hadronic collisions~\cite{Bzdak:2013zma,Schenke:2014zha,Albacete:2016pmp,Albacete:2016gxu,Mantysaari:2017cni,Schlichting:2014ipa}. This is especially true at high energies, where a significant fraction of the hadron’s momentum is carried by small-$x$ gluons, leading to a rapid growth in gluon density. In this high-density regime, gluon recombination and other non-linear QCD effects are expected to play an important role~\cite{Garcia-Montero:2025hys}. 

High-energy deep inelastic scattering (DIS) processes, found in electron-hadron collisions and in ultra-peripheral collisions (UPC) at the Large Hadron Collider (LHC) and the Relativistic Heavy-Ion Collider (RHIC)~\cite{Klein:2020nvu,Baltz:2007kq}, offer a theoretically robust framework for investigating the internal partonic structure of hadronic states. Furthermore, due to the intrinsic sensitivity of DIS processes to low-$x$ degrees of freedom in the target, these processes let us probe the transverse distributions to gain further understanding of the behavior of quarks and gluons in hadronic bound states in different energy regimes.

 In exclusive vector meson production, no net color charge transfer occurs between the dipole probe and the target hadron. Thus, the only correlations in transverse momentum and rapidity between the vector meson and the outgoing scattering products, or decay products of the dissociated target, are due to direct momentum transfer, which probes the color-charge distributions inside the hadron~\cite{Good:1960ba}.

In the asymptotic high-energy regime, this interaction is most effectively described within the dipole formalism~\cite{Kovchegov:2012mbw}, wherein the electron radiates a virtual photon, that subsequently undergoes quantum fluctuations into a quark-antiquark (\qqbar) pair.
 
A computational advantage of the dipole picture is the separation of the whole process into three distinct and statistically independent sub-processes: An electron initially emits a virtual photon,
which dissociates under quark-antiquark pair creation with a certain probability. This probability depends on the correlation of the wave functions of the initial and final states (photon and \qqbar~pair respectively), which in turn depend on the virtuality and polarization of the photon and the resulting dipole. The color dipole then interacts with the color charges present in the proton, transferring energy and momentum to the target. This interaction is usually represented as a two-gluon ladder in perturbative QCD. In scattering experiments, the decay products of the vector meson are detected and their energy and momentum analyzed~\cite{H1:2013okq}. The target proton either remains intact in an entirely elastic scattering (coherent scattering) or dissociates (incoherent scattering) due to the energy imparted to it.

In this work, we investigate saturation effects in the proton using the dipole picture of deep inelastic scattering. We  model the interaction between the dipole and the target with the Color Glass Condensate (CGC), an effective field theory of QCD that naturally incorporates multiple scattering and coherence effects, which become increasingly important at high gluon densities~\cite{Gelis:2010nm}. We compute both the coherent and incoherent cross sections for exclusive vector meson production using the fully non-relativistic expression for the overlap function, and compare our results to the analytically tractable dilute (leading-twist) limit~\cite{Demirci:2021kya,Demirci:2022wuy}. Following the methodology of previous studies, we exploit the hierarchy of relevant scales to disentangle the different sources of fluctuations, distinguishing between color charge fluctuations and geometric fluctuations of the proton structure, included phenomenologically successful hotspot models of the proton structure~\cite{Eremin:2003qn,PHENIX:2013ehw,Schlichting:2014ipa,Mantysaari:2016ykx,Albacete:2016pmp}. This framework allows us to systematically explore not only the impact on the coherent cross section, but also how high gluon densities affect different sources of incoherent fluctuations.

This paper is organized as follows: In \cref{sec:setup}, we introduce the theoretical framework, and we detail the dipole picture for modeling exclusive vector meson production in deep inelastic scattering, outline the derivation of coherent and incoherent cross sections, and describe the Color Glass Condensate framework combined with a hotspot-based proton model. \cref{sec:results} presents our numerical results, exploring the effects of saturation on both the coherent and incoherent vector-meson-production cross sections, with particular attention to the dependencies on color-charge density and mass of the heavy quarks that form the vector meson. Comparisons to experimental data are discussed to assess the effectiveness of the model. In \cref{sec:conclusion}, we conclude the paper by summarizing key findings and pointing out possible future research directions. Additionally, in Appendix~\ref{app:correlations}, the reader can find detailed derivations of the dipole cross section and the color correlation functions for both dense and dilute limits. Appendix~\ref{app:numerics} explains the numerical integration methods used and the specifics of hotspot sampling.

\section{Theoretical framework} \label{sec:setup}

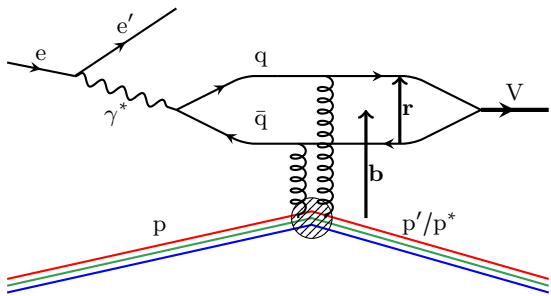
\begin{figure}[t]
  \begin{center}
    \begin{tikzpicture}[scale=0.9,
        fermion/.style={thick},
        photon/.style={decorate, decoration={snake, amplitude=2pt, segment length=8pt}, thick},
        gluon/.style={decorate, decoration={coil, aspect=0.6, segment length=0.15cm, amplitude=0.1cm}, thick},
        midarrow/.style={postaction={decorate, decoration={markings, mark=at position .5 with {\arrow{stealth}}}}},
      ]
        \draw[fermion, midarrow] (0,-0.3) -- (1,-0.5) node[midway, above] {$\mathrm{e}$};
        \draw[fermion, midarrow] (1,-0.5) -- (2.5,0.5) node[midway, above] {$\mathrm{e'}$};
        \draw[photon] (1,-0.5) -- (2.5,-1.0) node[midway, below left=0pt and -5pt] {$\gamma^*$};

        \draw[fermion, midarrow, rounded corners] (2.5,-1.0) -- (3.5,-0.5) -- (4.0,-0.5) node[midway, above] {$\mathrm{q}$} node[midway, below=10pt] {$\bar{\mathrm{q}}$};
        \draw[fermion, midarrow, rounded corners] (4.0,-0.5) -- (6.0,-0.5) -- (7.0,-1.0);
        \draw[fermion, midarrow, rounded corners] (4.0,-1.5) -- (3.5,-1.5) -- (2.5,-1.0);
        \draw[fermion, midarrow, rounded corners] (7.0,-1.0) -- (6.0,-1.5) -- (4.0,-1.5);

        \draw[ultra thick, midarrow] (7.0,-1.0) -- (8.0,-1.0) node[midway, above] {$\mathrm{V}$};

        \draw[color=red, fermion, shift={(0,0)}] (0,-3.5) -- node[midway, above, text=.] {$\mathrm{p}$} (4.5,-2.5) -- (8,-3.5) node[midway, above, text=.] {$\mathrm{p'/p^*}$};
        \draw[color=darkgreen, fermion, shift={(0,-0.1)}] (0,-3.5) -- (4.5,-2.5) -- (8,-3.5);
        \draw[color=blue, fermion, shift={(0,-0.2)}] (0,-3.5) -- (4.5,-2.5) -- (8,-3.5);

        \draw[pattern=north east lines, pattern color=.] (4.5,-2.6) circle (0.3);

        \draw[->, very thick, color=black] (5.8,-1.5) -- (5.8,-0.5) node[midway, right, shift={(-0.1,0.0)}] {$\rt$};
        \draw[->, very thick, color=black] (5.3,-2.6) -- (5.3,-1.0) node[midway, right, shift={(-0.1,-0.12)}] {$\bt$};
        \draw[gluon] (4.3,-1.5) -- (4.3,-2.6);
        \draw[gluon] (4.7,-0.5) -- (4.7,-2.6);
      \end{tikzpicture}
    \end{center}
    \caption{The process of exclusive vector meson production. Displayed are the incoming interaction particles (electron, which scatters off the virtual photon $\gamma^*$, with virtuality $Q^2$, and the proton) and the outgoing products (the vector meson and the proton p' or dissociated proton p$^*$).}
    \label{fig:evm}
\end{figure}

\subsection{Cross sections and scattering amplitude of the process}
In this work, we compute cross sections for the fully diffractive (coherent) process, $\gamma^* \mathrm{p} \rightarrow \mathrm{V} \mathrm{p}$, and the incoherent process, $\gamma^* \mathrm{p} \rightarrow \mathrm{V} \mathrm{p}^*$. In the Good-Walker picture~\cite{Good:1960ba}, the coherent cross section is given by the square of the average of the scattering amplitude as~\cite{Kowalski:2006hc}
\begin{equation}
	 \frac{\rmd\sigma\up{\rm coh}_\mathrm{T,L}}{\rmd t} = \frac{1}{16\pi} \bigabs{ \bigavg{\bigavg{A\up{\gamma^* p \rightarrow V p}\down{T,L}}} }^2 
	\label{SigmaCoherent}
\end{equation}
and the incoherent cross section computes via the variance of the amplitude:
\begin{equation}
	\frac{\rmd \sigma\up{\rm inc}\down{T,L}}{\rmd t}  = \frac{1}{16\pi} \Bigl( \bigavg{\bigavg{\bigabs{A\up{\gamma^* p \longrightarrow V p}\down{T,L}}^2}} - \bigabs{ \bigavg{\bigavg{A\up{\gamma^* p \longrightarrow V p}\down{T,L}}}}^2  \Bigr)\,, 
	\label{SigmaIncoherent}
\end{equation}
where $t = -\bfDelta^2$ is the Mandelstam variable describing momentum transfer from the dipole to the proton. The double averaging process is introduced here since, in what follows, we will include two different sources of fluctuations in our computations. On one hand, color fluctuations in the target will arise at scales $Q\down{s}^{-1}$, the saturation scale of the target. On the other hand, we include geometrical fluctuations, which arise at scales similar to the nucleon radii. The double average, which we define with the compact notation of
\begin{equation}
  \langle\langle O \rangle\rangle \equiv \langle\langle O \rangle\down{c}\rangle\down{h}\,,
	\label{eq:double_average}
\end{equation}
denotes averaging of color fluctuations in the target, labeled as $\mathrm{c}$, and the averaging of event-by-event geometrical fluctuations, labeled as $\mathrm{h}$.
 
The ZEUS collaboration has traditionally defined the total cross section as the direct sum of the cross sections from virtual photons with transverse and longitudinal polarizations~\cite{Kowalski:2006hc}
\begin{align}
 	\frac{\rmd \sigma}{\rmd t} \approx \frac{\rmd \sigma\down{T}}{\rmd t} + \frac{\rmd \sigma\down{L}}{\rmd t} \,.
\end{align}

We will henceforth not show the subscripts $\mathrm{T}$ or $\mathrm{L}$, and the reader should consider all given cross sections to be the sum over polarizations, unless stated otherwise.

To better extract how the incoherent production process is affected by the fluctuations in the initial wavefunction of the proton target, we can separate the $\rmd\sigma\up{inc}/\rmd t$ into two different parts as

\begin{align}
	\frac{\rmd \sigma\down{inc}}{\rmd t} & = \underbrace{\frac{1}{16\pi} \bigavg{ \avg{\abs{A}^2}\down{c}-\abs{\avg{A}\down{c}}^2 }\down{h}}_{\mathrm{color~fluc.}}
	\nn \\ & \quad + \underbrace{\frac{1}{16\pi} \bigavg{\abs{\avg{A}\down{c}}^2}\down{h}-\bigabs{\bigavg{\avg{A}\down{c}}\down{h}}^2}_{\mathrm{hotspot~fluc.}}\,,
\end{align}
which clearly shows that color fluctuations are given by the hotspot average of the color variance, while the hotspot fluctuations are calculated as the hotspot variance of color averaged amplitudes.

\subsection{Scattering amplitude in the dipole picture}
In the dipole picture of DIS, the scattering amplitude for exclusive vector meson production can be split into three distinct sub-processes. The incoming photon splits into a \qqbar~pair, which interacts with the target and is eventually projected into a meson state. The amplitude for this tripartite process can then be given, at leading order~\cite{Mantysaari:2020axf,Kowalski:2006hc}, by
\begin{align}
    A\up{\gamma^* p \rightarrow V p}\down{T,L}& (Q^2, \bfDelta) =\rmi \int \rmd^2 \rt~ \int \rmd^2 \bt~ \int \dfrac{\rmd z}{4\pi} \, \frac{\rmd \sigma\up{p}\down{dip}(\rt)}{\rmd^2\bt}\nn \\\
      &  \quad\quad \times \rme^{-\rmi \left[\bt + \left( \tfrac{1}{2}-z \right) \rt \right] \bfDelta }
         \left( \Psi_\gamma^* \Psi\down{V} \right)\down{T,L} (Q^2,\rt,z)\,,
    \label{eq:amplitude}
\end{align}
where the properties of the probe itself are given by $\left( \Psi_\gamma^* \Psi\down{V} \right)\down{T,L}$, the wave-function overlap for the virtual photon splitting into a \qqbar~and subsequent vector meson production. On the other hand, the interaction of the \qqbar~pair with the strongly interacting target is described by the dipole cross section, $\frac{\rmd\sigma\up{p}\down{dip}(\rt)}{\rmd^2\bt}$, where  $\rt$ denotes the transverse dipole size and $\bt$ the impact parameter of the dipole relative to the target's center of mass,  as depicted in Fig.~\ref{fig:evm}. These two characteristic sizes are related to the transverse positions of the quark and antiquark in the dipole ($\xt$ and $\yt$, respectively), by
\begin{align}
	\bt = \frac{\xt+\yt}{2}\quad \text{and} \quad \rt = \xt - \yt\,.
\end{align}

The photon virtuality is given by $Q^2$, and $z$ is the fraction of the light-cone momentum of the incoming photon that is carried by the quark in the dipole, while naturally the antiquark carries $1-z$. The total transverse momentum transfer into the target is described by $\bfDelta$.

In the following, we will focus on the production of vector mesons from heavier quark species ($\mathrm{c}$ and $\mathrm{b}$).
Since the primary objective of this study is to investigate the role of saturation effects, we follow previous works~\cite{Demirci:2022wuy,Kowalski:2006hc,Lappi:2020ufv} and employ the non-relativistic limit of the vector meson wave functions, which results in the overlaps
\begin{align}
  \left( \Psi_\gamma^* \Psi\down{V} \right)\down{T} &= -A\down{q} \sqrt{2 m\down{q} N\down{c}} e\down{q} e K_0(\varepsilon \abs{\rt}) \delta \left( z - \tfrac{1}{2} \right)
  \label{eq:psipsiT}
\end{align}
for photons with transverse polarization, and
\begin{align}
  \left( \Psi_\gamma^* \Psi\down{V} \right)\down{L} &= -2 A\down{q} \sqrt{\tfrac{2 N\down{c}}{m\down{q}}} e\down{q} e Q z \left( 1-z \right) K_0(\varepsilon \abs{\rt}) \delta \left( z - \tfrac{1}{2} \right)\,.
     \label{eq:psipsiL}
\end{align}
for photons with longitudinal polarization.

While \cref{eq:psipsiT,eq:psipsiL} can be seen as the leading terms in an expansion of the relative velocities of the (heavy) quark-antiquark pair, it should be noted that terms sub-leading in power can also have significant effect on the cross section, in particular for charm quarks (see e.g.~\cite{Mantysaari:2025idf}).

In \cref{eq:psipsiT,eq:psipsiL}, $A\down{q}$ is a normalization constant controlling the total decay width of the respective vector meson to electron-positron pair via~\cite{Lappi:2020ufv}
\begin{align}
    \Gamma(\mathrm{q}\ob{\mathrm{q}} \rightarrow \mathrm{e}^+\mathrm{e}^-) = A\down{q}^2 \frac{4\pi e\down{q}^2 \alpha\down{em}}{m\down{q}^2}\,,
\end{align}
where $e=\sqrt{4\pi \alpha\down{em}}$ is the electromagnetic coupling constant, and $e\down{q}$ describes fractional electric charge of the quark. The governing quantity for the width of the wave-function overlap is~\cite{Lappi:2020ufv}
\begin{align}
    \varepsilon = \sqrt{Q^2 z(1-z) + m\down{q}^2}\,.
\end{align}

In the non-relativistic limit, the produced quarks both carry the same momentum and energy, making the impact parameter, $\bt$, the sole Fourier conjugate to the momentum transfer $\bfDelta$
\begin{align}
	A\up{\gamma^* p \rightarrow V p}\down{T,L} & (Q^2, \bfDelta) = \frac{\rmi}{4\pi} \int \rmd^2 \rt \int \rmd^2 \bt~ \rme^{-\rmi \bt \bfDelta} \frac{\rmd\sigma\up{p}\down{dip}(\rt)}{\rmd^2\bt} \nonumber\\
	&\qquad \times\left( \Psi_\gamma^* \Psi\down{V} \right)\down{T,L} (Q^2,\rt,\tfrac{1}{2})\,.
\end{align}
which significantly simplifies in the computation of the amplitude in the non-relativistic limit.

It is important to note that the energy transfer $\bfDelta$ in the scattering amplitude also includes information about the direction of transfer. When geometrical fluctuations are taken into account in the proton, the individual events become anisotropic. As experimental data on the \mbox{\textnormal(in-)coherent} cross sections is reported in terms of the variable $t$ and insensitive to these fluctuations, one thus needs to average over the transverse angle of $\bfDelta$ to compare to data. This procedure is explained in Appendix~\ref{app:phi-avg}.

\subsection{Dipole cross section and correlations }
All information about the interaction of the \qqbar-pair~with the target is contained in the dipole cross section,
\begin{align}
    \frac{\rmd\sigma\up{p}\down{dip}}{\rmd^2\bt} (\bt,\rt) = 2 \left(1 - D(\bt,\rt)\right)\,,
\end{align}
which, for a single configuration of color charges, can be written using the dipole correlator $D$, namely,
\begin{align}
    D (\bt,\rt) = \frac{1}{\Nc}\mathrm{tr} \left[ V\left(\bt + \tfrac{\rt}{2}\right) V^\dagger\left(\bt - \tfrac{\rt}{2}\right) \right]\,.
    \label{eq:dipole_op}
\end{align}
In this trace, $V(\xt)$ corresponds to a light-like Wilson line operator, which describes the interaction of the passing quark with the target color fields. It is explicitly written as a resummation of an infinite number of insertions of classical color fields:
\begin{align}
    V (\xt) = P_+  \exp & \left\{\rmi g \int^\infty_{-\infty} \rmd z^+ A^-_a (z^+,\xt)t^a \right\}
    \label{eq:wilson}
\end{align}
Integration along $z^+$, which is the light-cone coordinate of a source moving along the longitudinal axis of collision, describes the quarks passing through the target.
Additionally, the gauge field  $A_a^-$ and its fundamental generator $t^a$ constitute the gluon field at transverse position $\xt$. In the CGC formalism~\cite{McLerran:1993ni,McLerran:1993ka} these fields are generated in response to the  large-$x$ color sources in the target, $\rho^a$, and are explicitly given by 
\begin{equation}
	A^-_a (z^+,\xt)=\int \rmd^2 \mathbf{z} \, G(\xt - \mathbf{z}) \rho_a (z^+,\mathbf{z})\,.
\end{equation} 

The function $G(\xt-\zt)$ is the Green's function associated with the propagation of a free tranverse gluon, namely~\cite{McLerran:1994vd,Kovner:1995ts}
\begin{align}
    G(\xt-\zt) = \int \frac{\rmd^2\kt}{(2\pi)^2} \frac{\rme^{\rmi\kt(\xt-\zt)}}{\kt^2+m^2}\,,
    \label{eq:GreensFunReg}
\end{align}
where $m$ has been introduced as an IR-regulator.

We will compute the vector-meson production cross section in two limits, namely for a dense and for a dilute target. When considering a dense target, the dipole cross section is taken to be the trace of the full Wilson line, while considering a dilute target (also commonly known as leading-twist approximation) corresponds to an expansion of the dipole operator only up to the first non-trivial order in the gluon fields (and thus, the color-charge density). The leading-twist vector meson cross section was studied in previous work~\cite{Demirci:2022wuy}.
We note that in both cases,  we need the color average, i.e. the average over color-charge configurations, of the dipole cross section $D$ 
and its two-point correlation function $DD$ to evaluate coherent and incoherent cross-sections. These color averages

can be performed analytically, as for a static set of color sources (no $x$ evolution), the dipole cross section and its two-point function depend only on the logarithm of the dipole correlation function \cite{Gelis:2001da,Lappi:2017skr,Deganutti:2023qct}
\begin{align}
    G_{\xt\yt} = \log\left[1-\frac{1}{2}\left\langle \rmd\sigma\up{p}\down{dip}/\rmd^2\bt\right\rangle\down{c}\right]
\end{align}
The function $G_{\xt\yt}$ can be obtained in a diverse array of models, with popular choices including the Golec-Biernat-Wüsthoff (GBW) model \cite{Golec-Biernat:1999qor}, the McLerran-Venugopalan (MV) model~\cite{McLerran:1993ni,McLerran:1993ka} and the IP-Sat model~\cite{Rezaeian:2013eia,Kowalski:2003hm}. We discuss the calculation of $G_{\xt\yt}$ in our hot-spot model further below.\\

\noindent \textbf{Dense target:}  The dense limit allows for infinitely many scatterings of the \qqbar~pair off the target. For the case of the color average of the dipole operator, the resulting function is simply given by
\begin{align}
    \bigggavg{\frac{\rmd\sigma\up{p,dense}\down{dip}}{\rmd^2\bt}(\bt, \rt)} = 2 \left( 1 - e^{G_{\xt\yt}} \right)
    \label{eq:dsigmad2b-dense}
\end{align}
Similarly, the color-averaged dipole-dipole color correlator can be expressed as~\cite{Blaizot:2004wv,Dominguez:2011wm}
\begin{align}
    &\biggggavg{ \frac{\rmd\sigma\up{p,dense}\down{dip}}{\rmd^2\bt}(\bt,\rt) \frac{\rmd\sigma\up{p,dense}\down{dip}}{\rmd^2\ob{\bt}}(\ob{\bt},\ob{\rt}) }  \\
    & = 4 \left( 1 - \rme^{G_{\xt\yt}} - \rme^{G_{\xbt\ybt}} + \frac{\rme^{\frac{a+d}{2}}}{2}  \left[(1+F) \rme^{\frac{1}{2}f} + (1-F) \rme^{-\frac{1}{2}f} \right] \right)\nn\,,
    \label{eq:dipole_squared}
\end{align}
and we include the derivation of this result in Appendix~\ref{app:correlations}. The functions to be evaluated for this correlation are
\begin{equation}
	    f = \sqrt{4bc+(a-d)^2} \quad \text{and} \quad
	F = \frac{a-d+\frac{2b}{\Nc}}{f} \,,
\end{equation}
where the short-hands $a$ through $d$ are defined as follows:
\begin{equation}
	\begin{split}
		    a &= G_{\xt\yt} + G_{\xbt\ybt} - \frac{1}{\Nc^2-1}T_{\xt\yt,\xbt\ybt} \\
		b &= \frac{1}{2\CF}T_{\xt\yt,\xbt\ybt} \quad \quad c = \frac{1}{2\CF}T_{\xt\ybt,\xbt\yt} \\
		d &= G_{\xt\ybt} + G_{\xbt\yt} - \frac{1}{\Nc^2-1}T_{\xt\ybt,\xbt\yt}
	\end{split}
\end{equation}
Additionally, $T$ denotes the gluon-exchange function, which can be written in terms of $G_{\xt\yt}$ as 
\begin{align}
T_{\xt_1\xt_2,\xt_3,\xt_4} = G_{\xt_1\xt_4} + G_{\xt_2\xt_3} - G_{\xt_1\xt_3} - G_{\xt_2\xt_4}\,.
\end{align}

\noindent \textbf{Dilute target:}
Expanding the Wilson line operators to second order in the color-charge density yields the dipole cross section
\begin{align}
    \bigggavg{\frac{\rmd\sigma\up{p,dilute}\down{dip}}{\rmd^2\bt}(\bt, \rt)} \approx -2G_{\xt\yt}
    \label{eq:dsigmad2b-dilute}
\end{align}
and the dipole-dipole cross section, which appears in the incoherent cross section \cite{Lappi:2015vta}
\begin{align}
    &\biggggavg{ \frac{\rmd\sigma\up{p,dilute}\down{dip}}{\rmd^2\bt}(\bt,\rt) \frac{\rmd\sigma\up{p,dilute}\down{dip}}{\rmd^2\ob{\bt}}(\ob{\bt},\ob{\rt}) }  \\
    & \approx 4 \left( G_{\xt \yt}G_{\xbt \ybt} + \frac{1}{2(\Nc^2-1)} \left( G_{\xt \xbt} + G_{\yt \ybt} - G_{\xt \ybt} - G_{\yt \xbt} \right)^2 \right).\nn
\end{align}
where for completeness, the computation of these leading-twist approximations can be found in Appendix~\ref{app:correlations}.

\subsection{Proton hotspot model}

We will use the hotspot model~\cite{Albacete:2016pmp,Albacete:2017joc} to quantify position-space (geometrical) fluctuations in the proton. In this model, the proton presents event-by-event fluctuations of the color charges, effectively accumulating over $N\down{h}$ dense areas, hence the name of the model. Using the hotspot model, the measured observables undergo averaging as calculated within a double average: first over the positions $\bt_i$ of the $N\down{h}$ hotspots, and then over the configurations of color charges ($\rho_a(\xt)$) within the hotspots.

For the distribution of the positions of such hotspots, we will assume a Gaussian distribution, which can be expressed in terms of the thickness function of the proton
\begin{align}
  T\down{p}(\xt) = \frac{1}{2\pi R^2} \exp \left\{ -\frac{\xt^2}{2R^2} \right\}
	\label{eq:thickness}
\end{align}
which is normalized to unity. We can then explicitly define the double average from \cref{eq:double_average} as
\begin{align}
	\avg{\avg{O}\down{c}}\down{h} \equiv& \left( \frac{2\pi R^2}{\NH} \right) \int \prod_{i=1}^{\NH} \left[ \rmd^2\bt_i~ T(\bt_i - \textbf{B}) \right] \nn \\
	&\quad\times\delta^{(2)} \left( \frac{1}{\NH} \sum_{i=1}^{\NH} \bt_i - \textbf{B} \right) \avg{O}\down{c}\,,
\end{align}
where the 2d $\delta$ constrains the distribution in such a manner that the positional average of all hotspots is in the center of the proton, $\mathbf{B}$. For the case of $\mathrm{e}\mathrm{p}$ collisions, we can set $\mathbf{B} = 0$ without loss of generality.
\begin{figure}[t]
	\centering
	
    \begin{tikzpicture}[scale=0.7, shift={(3.0,2.0)}]
      \tikzfading[name=circlefade, inner color=transparent!30, outer color=transparent!100]

      \draw[very thick, dashed] (0,0) circle (2.57);

      \draw[very thick, dotted, color=blue] (-1.0,0.0) circle (1.18);
      \fill[blue, path fading=circlefade, fading transform={shift={(0,0)}}] (-1.0,0.0) circle (1.18*1.5);
      \draw[very thick, dotted, color=darkgreen] (0.5,-0.866) circle (1.18);
      \fill[darkgreen, path fading=circlefade, fading transform={shift={(0,0)}}] (0.5,-0.866) circle (1.18*1.5);
      \draw[very thick, dotted, color=red] (0.5,0.866) circle (1.18);
      \fill[red, path fading=circlefade, fading transform={shift={(0,0)}}] (0.5,0.866) circle (1.18*1.5);

      \draw[color=black, very thick, <->] (0,0) -- (2.57,0) node[midway, above] {$R$};
      \draw[color=black, very thick, <->] (0.5,0.866) -- (1.68,0.866) node[midway, above] {$r_\mathrm{H}$};

      \draw[color=black, very thick, ->] (0.0,0.0) -- (1.2,-1.4) node[midway, above, shift={(0.2,-0.2)}] {$\bt$};
      \draw[color=black, very thick, ->] (1.9,-1.1) -- (0.5,-1.7) node[midway, below, shift={(0.0,0.0)}] {$\rt$};
      \draw[mark=x, mark size=4, very thick, color=black] plot coordinates {(0.5,-1.7)} node[left] {$\xt$};
      \draw[mark=x, mark size=4, very thick, color=black] plot coordinates {(1.9,-1.1)} node[right] {$\yt$};
    \end{tikzpicture}

	\caption{Illustration of the hotspot proton model, which shows an arbitrary configuration of $\NH=3$ hotspots. The hotspots are not confined to the proton and radiate some color charge beyond the size of the proton. The passing dipole with quarks at positions $\xt$ and $\yt$ is also drawn.}
	\label{fig:hotspot-model}
\end{figure}
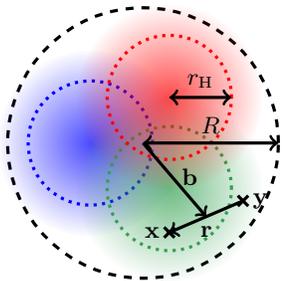

With regards to the averages over color charge distributions, we will use the MV model, which assumes Gaussian correlations between the color charges, such that
\begin{align}
  \avg{\rho_a(\xt)\rho_b(\yt)}\down{c} = \mu\down{p}^{2}\Big(\frac{\xt+\yt}{2}\Big) \delta^{(2)} \left(\xt-\yt\right) \delta^{ab}\,,
\end{align}
while the one-point function $\avg{\rho_a(\xt)}\down{c} = 0$ vanishes to guarantee color neutrality in the target proton~\cite{Albacete:2016pmp}. Here $\mu\down{p}^2$ describes the color-charge distribution in the transverse plane, which we take as the the sum of the charge profiles of $\NH$ hotspots  located in positions $\bt_i$, where $i \in \{1, ..., \NH\}$, as
\begin{align}
\mu\down{p}^{2}(\xt)=\sum_{i=1}^{\NH} \mu\down{h}^2\left(\xt - \bt_i\right)\,.
\end{align}
Event-by-event spatial distributions in the transverse plane are modelled by assuming a Gaussian profile for the color charge density of each hotspot, namely
\begin{align}
	\mu^2\down{h}(\xt) = \frac{\mu_0^2}{2\pi r\down{h}^2} \exp \left\{ -\frac{\xt^2}{2 r\down{h}^2} \right\} ,
	\label{eq:hotspot}
\end{align}
where the effective width of these distributions is $r\down{h}$.  The dimensionless parameter $\mu_0$ determines the total normalization of color charge present in the proton target and $R$ its effective radial width. Generally, we expect the color-charge normalization parameter, $\mu_0$,  to be dependent on the center-of-mass energy of the process,  as a higher $\sqrt{s}$ will induce greater charge densities in the target. 

We also note in passing, that the double average procedure induces a non-trivial color charge density distribution on average. In particular, at the leading-twist limit, the emerging distribution can be understood at the analytical level, where the induced density is given by  $\avg{\avg{\rho_a(\xt)\rho_b(\yt)}\down{c}}\down{h} =\delta_{ab}\delta^{(2)}(\xt-\yt) \bar{\mu}^2(\bt-\mathbf{B})$. The new  effective color correlation, also Gaussian, is explicitly given by~\cite{Demirci:2022wuy}
\begin{equation}
\bar{\mu}^2(\bt-\mathbf{B}) =\frac{\mu_0^2}{2\pi R\down{C}^2} \exp\left\{-\frac{(\bt-\mathbf{B})^2}{2 R\down{C}^2}\right\}\,, 
\end{equation} 
where the \emph{coherent radius}, $R\down{C}= r\down{h} + R^2 (N\down{h}-1)/N\down{h}$, corresponds to the new effective width of the proton due to the hotspot average~\cite{Demirci:2022wuy}.

Based on the hot-spot model, we can now explicitly compute the correlation function $G_{\xt\yt}$. Since, in accordance with the discussion in Appendix~\ref{app:correlations}, the correlation function $G_{\xt\yt}$ is linear in the color-charge density correlator $\rho_{a}\rho_{b}$, individual hot-spots contribute additively to $G_{\xt\yt}$ and it is instructive to first consider the contribution of a single hot-spot located at a position $\bt=0$. We can then express $G_{\xt\yt}$ semi-analytically (see Appendix~\ref{app:correlations}), which is given by
\begin{widetext}
	\begin{equation}
		\begin{split}
			G_{\xt \yt,r\down{h}}= \frac{g^2\mu_0^2\CF}{(4\pi)^2}&\int_0^\infty \rmd u~ \int_0^\infty \rmd v~ \frac{\rme^{-m^2(u+v)}}{uv+\frac{r\down{h}^2}{2}(u+v)}  \quad \Biggl[ \exp \left\{ -\frac{1}{4}\frac{u\,\yt^2+v\,\xt^2+\frac{r\down{h}^2}{2}(\xt-\yt)^2}{uv+\frac{r\down{h}^2}{2}(u+v)} \right\}  \\
			& -\frac{1}{2}\exp \left\{ -\frac{1}{4} \frac{(u+v)\,\xt^2}{uv+\frac{r\down{h}^2}{2}(u+v)} \right\}  -\frac{1}{2}\exp \left\{ -\frac{1}{4} \frac{(u+v)\,\yt^2}{uv+\frac{r\down{h}^2}{2}(u+v)} \right\} \Biggr]\,,
		\end{split}
		\label{eq:GXYR}
	\end{equation}
\end{widetext}
where $\CF=\frac{\Nc^2-1}{2\Nc}$ is the fundamental Casimir and $m$ is the IR-regulator introduced in Eq.~(\ref{eq:GreensFunReg}).

By considering the contributions from all hotspots, the  total function $G_{\xt\yt}$ can then simply be computed using
\begin{equation}
	\begin{split}
    G_{\xt \yt} = &\sum_{i=1}^{\NH} G_{\xt-\bt_i, \yt-\bt_i, r\down{h}}\,,
	\end{split}
    \label{eq:sumGxy}
\end{equation}
where $ G_{\xt-\bt_i, \yt-\bt_i, r\down{h}}$ is given by shifting $G_{\xt\yt,r\down{h}}$, as defined by \cref{eq:GXYR}, to the center of the hotspot, $\mathbf{b}_i$.

Clearly, the formulation of a model in terms of microscopic charges, where Wilson lines are obtained by solving the YM field equations, has the advantage that it automatically ensures the consistency of all $N$-point functions. Notably, this is not the case for many phenomenological parametrizations of the dipole cross section, which are not based on microscopic considerations, but instead provide an explicit parametrization of the dipole cross-section $D$ directly as a function of $\rt$ and $\bt$. We found that inserting popular parametrizations such as the GBW model~\cite{Golec-Biernat:1998zce} into higher correlations of the dipole operator, \cref{eq:dipole_op}, leads to inconsistencies for higher-order $N$-point functions, where e.g. the absence of cancellatiosn in the quadrupole correlation leads to a non-vanishing constant even for $|\bt|\to\infty$, resulting in divergent amplitudes. For more information, the reader is referred to  Appendix~\ref{app:correlations}.

\section{Results} \label{sec:results}
We will now present numerical computations for coherent and incoherent production of $\mathrm{J}/\psi$ and $\Upsilon$ vector mesons. 
In order to evaluate the cross-sections, we perform a Monte-Carlo sampling of the hotspot positions, and subsequently evaluate the scattering amplitude by numerical integration using the \texttt{cubature} package~\cite{cubature}. The software for the calculation of scattering amplitudes for single hot-spot configurations and the necessary analysis of these results is available at~\cite{EIC-Dipper}, where the  reader can also find the data and plotting scripts for the results of this paper.

We set the decay width to $A\down{c} = 0.211\,\mathrm{GeV}^{3/2}$ which has been accurately determined from experiments involving charmed states~\cite{Lappi:2020ufv}. Due to the lack of an accurate measurement of $A\down{b}$, we will use $A\down{b}=A\down{c}$ for the production cross section comparisons involving bottom quarks. Some other numerical parameters used are the electromagnetic coupling, $e = \sqrt{4\pi \alpha\down{em}}$, where $\alpha\down{em}^{-1} = 137.036$, and $e\down{q}$, which describes the fractional electric charge of the respective quarks. The quark masses are fixed to $m\down{c}  = 1.275\,\mathrm{GeV}$ for the charm quark, and $m\down{b} = 4.18\,\mathrm{GeV}$ for the bottom quark.

The color-charge density in each hotspot is varied with the parameter $\mu_0^2$, which appears in the dipole correlation function $G_{\xt\yt}$ together with the square of the gluon coupling constant $g^2$. Therefore, $\mu_0$ always enters the cross section as the dimensionless combination $g^2 \mu_0^2$, which, in accordance with previous work~\cite{Demirci:2022wuy}, was chosen to be $g^2\mu_0^2 = 6.574$ as the best fit value for comparison with the HERA H1 results~\cite{H1:2003ksk,H1:2013okq}. This will be the value reported in this work, unless explicitly stated otherwise.

\begin{figure}[t]
	\centering
	\includegraphics[width=0.5\textwidth]{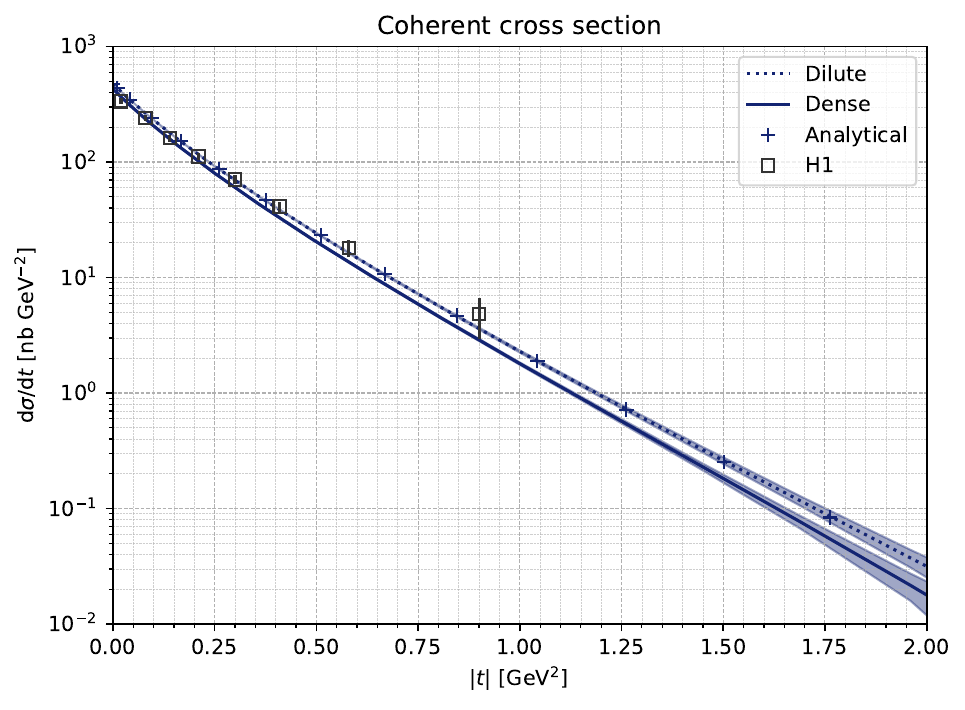}
	\caption{Coherent cross sections in the dense and dilute limits. Data labeled "Analytical" refers to results from previous work~\cite{Demirci:2021kya}, where the hotspot average was calculated analytically in the dilute limit and no numerical sampling was performed. H1 data is from~\cite{H1:2013okq}. The bands correspond to statistical error coming from the MC sampling of geometrical fluctuations in the proton.} 
	\label{fig:co}
\end{figure}

The photon virtuality is given by $Q^2 = 0.1\,\mathrm{GeV}^2$.
Additionally, the characteristic sizes of hotspot and proton were chosen as $r\down{h} = 0.165\,\mathrm{fm}$ and $R = 0.358\,\mathrm{fm}$ respectively.
We have also fixed the number of hotspots inside each proton to be $\NH=3$, for a coherent radius $R\down{C}=0.336\,\mathrm{fm}$. Finally, the IR-regulator in the Wilson line integral's Green's function is $m = 0.22\,\mathrm{GeV}$.

\begin{figure}[t]
\centering
\includegraphics[width=0.5\textwidth]{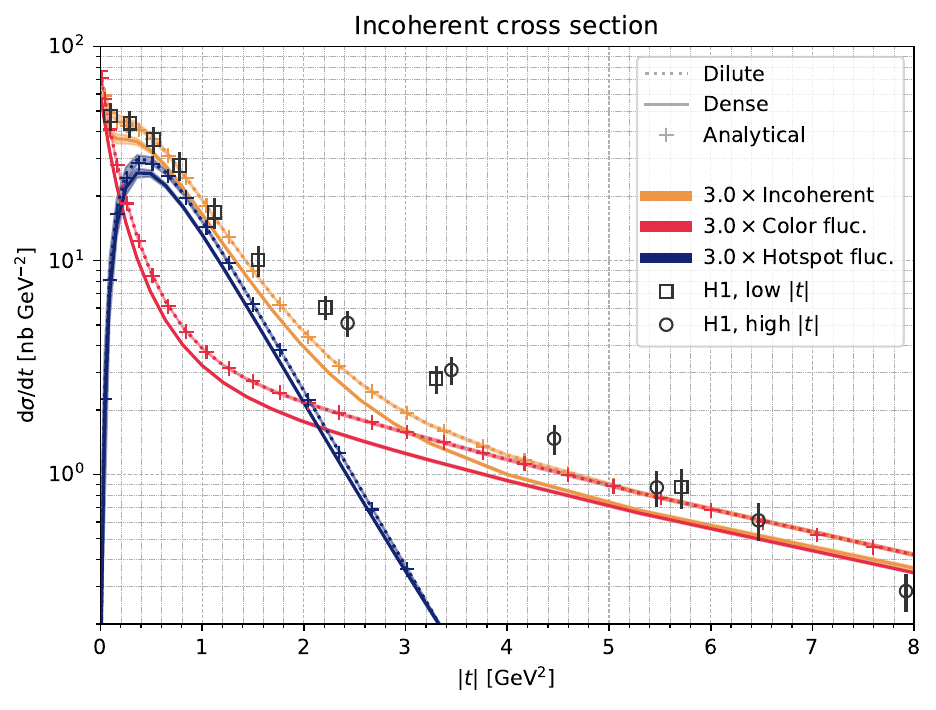}
 \caption{Incoherent cross sections in the dense and dilute limits. Also shown are the color and hotspot fluctuations making up the total cross section. H1 data is from~\cite{H1:2013okq} for small-$t$ and~\cite{H1:2003ksk} for large-$t$. The standard error is shown as bands.}
\label{fig:inco}
\end{figure}

The reader can find the coherent and incoherent cross sections as functions of $|t|=|\boldsymbol{\Delta}|^2$, in \cref{fig:co,fig:inco}, respectively. We compare the dilute and dense limits\footnote{We have validated our numerical integrator and sampling algorithms against the analytical results of Ref.~\cite{Demirci:2022wuy}, which are labeled "Analytical" in the figures. } for $\mathrm{J}/\psi$ production, and validate the chosen parameters against H1 data~\cite{H1:2003ksk,H1:2013okq}. In all the results we are presenting in this section, we average over 512 different hotspot configurations (events), where we use the same underlying sets of events for both the dilute and the dense limits. The bands shown correspond to the standard error in the statistical analysis of these events.

We observe a consistent suppression of the cross sections for the dense limit, in comparison to the dilute approximation. Naturally, this is expected from these models, as the dense limit contains scatterings to all orders. This means that the scattering quarks experience larger momentum broadening than in the single-scattering dilute limit. Computationally, one can understand this from the dipole correlation function, $G_{\xt\yt}$, which is non-positive for all positions $\xt$ and $\yt$. This results in the dense-target dipole cross section being consistently smaller than the dilute-target dipole cross section (compare \cref{eq:dsigmad2b-dense,eq:dsigmad2b-dilute}). In fact, it is also bounded from above, whereas the dilute dipole cross section is not. These effects ultimately lead to suppression of the scattering amplitude and thus cross sections. In comparison with the results presented in this paper for $\mathrm{e}\mathrm{p}$, we expect the suppression to be larger for the case of $\mathrm{e}\mathrm{A}$, since the effective color density will increase with the total number of nucleons.

As can be seen in \cref{fig:co}, the coherent cross sections for both the dense and dilute limits show exponential decrease with $\abs{t}$ for larger momentum transfers ($|t|>0.7\,\mathrm{GeV}^2$), where the slope is is affected by the size of the system~\cite{Demirci:2022wuy}. Hotspot fluctuations naturally change this slope by changing, event-by event, the effective color charge density and the effective size of the system. While this is very straightforward in the dilute case, leading to the coherent radius $R\down{C}$, in the case of a dense target, the new effective size tends to be more entangled to the remaining parameters. In \cref{fig:co}, this variation manifests in the error bands, which grow relative to the smaller cross sections at large $|t|$\footnote{The variation from the sampling for the coherent cross sections shows a relative increase with the transferred energy, and thus $\abs{t}$. This can be explained as follows: Firstly, the energy transfer is higher for dipoles that scatter closer to the center of the proton, where the average proton thickness is the highest. Secondly, the variation in hotspot positions induces variation of the effective thickness profiles event-by-event. This effect is also maximal in the center of the proton. The combination of these two effects then results in comparatively greater (statistical) variation for higher energy transfers.}.

\begin{figure}[t]
	\includegraphics[width=0.5\textwidth]{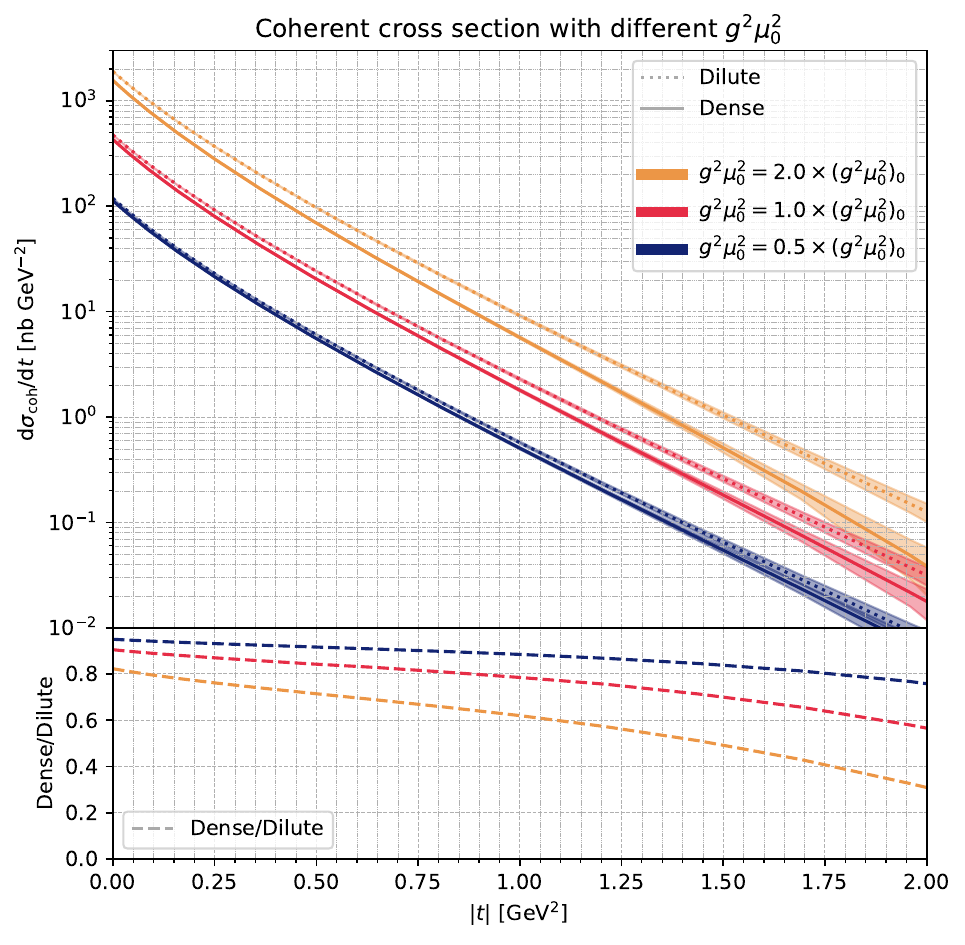}
	\caption{Coherent cross sections for different color-charge amounts $g^2\mu_0^2$. The standard error is shown as bands. The smaller diagram shows the ratio of dense to dilute model scattering cross sections. }
	\label{fig:co-g2mu02}
\end{figure}

\begin{figure}[t]
	\includegraphics[width=0.5\textwidth]{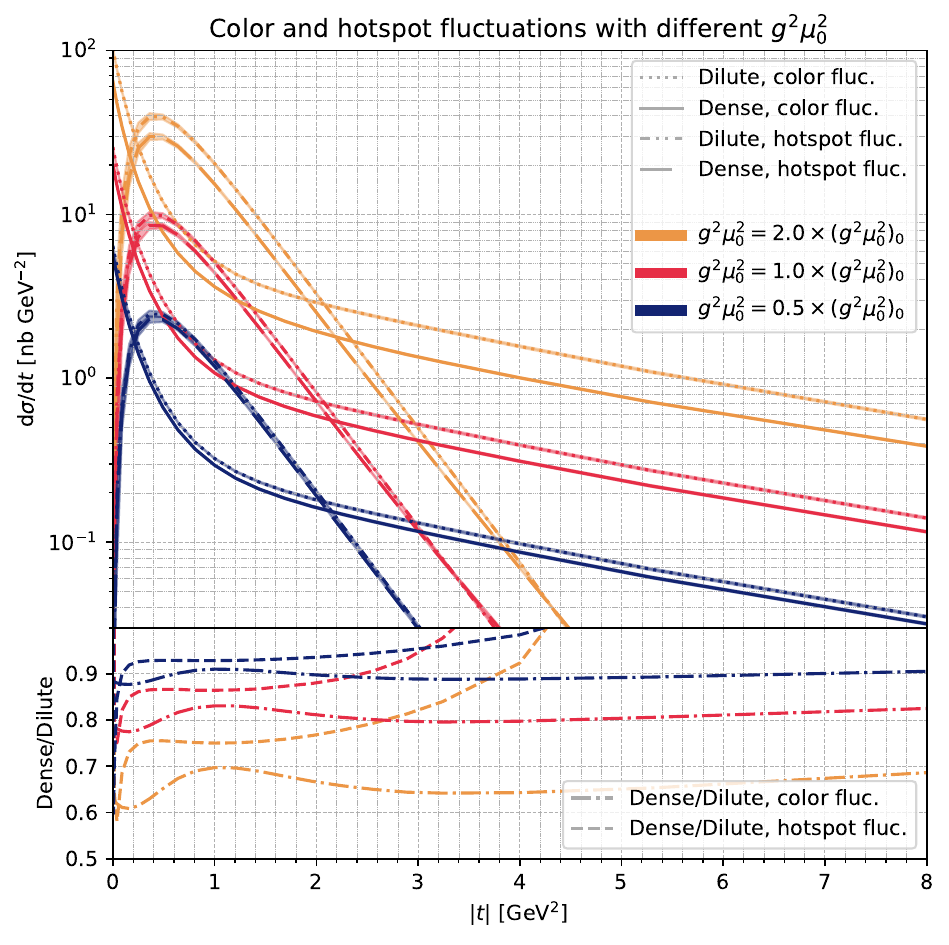}
	\caption{Color and hotspot fluctuations for different color-charge amounts $g^2\mu_0^2$. Shown are the color- and hotspot-fluctuation contributions making up the full incoherent cross section shown in \cref{fig:inco-g2mu02}. The standard error is shown as bands. The smaller diagram shows the ratio of dense to dilute model scattering cross sections.}
	\label{fig:inco-g2mu02-fluc}
\end{figure}

\begin{figure}[t]
	\includegraphics[width=0.5\textwidth]{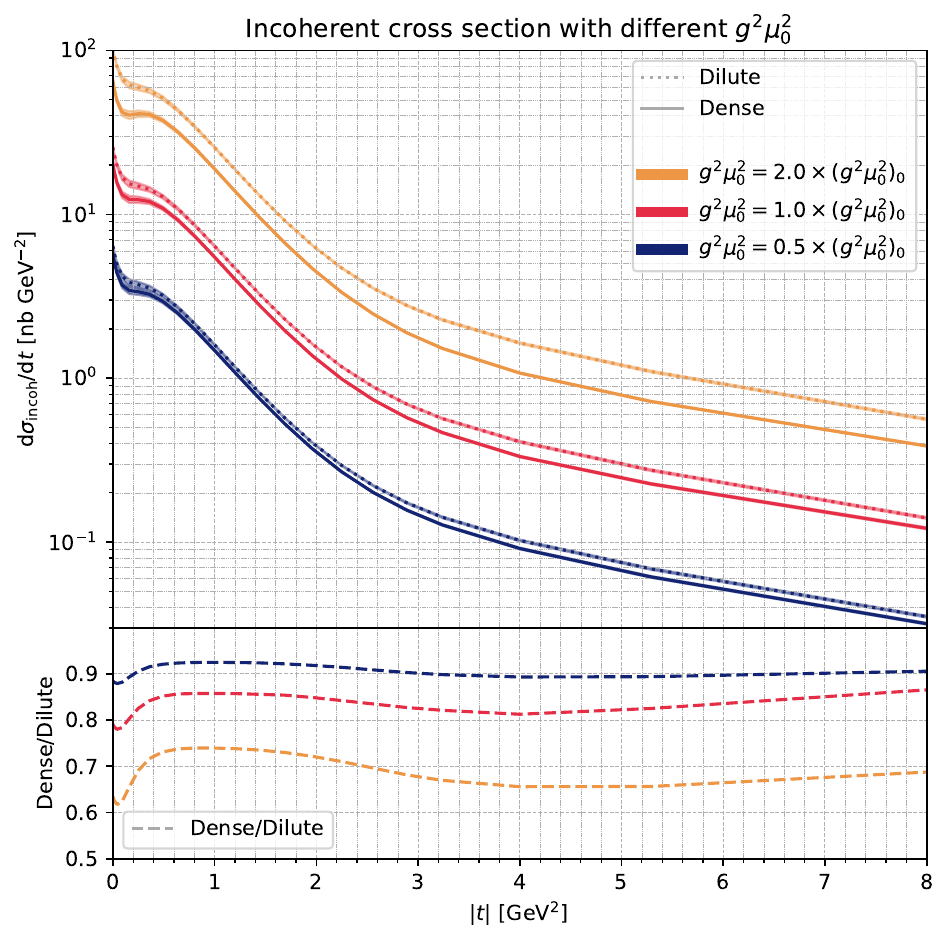}
	\caption{Incoherent cross section for different color-charge amounts $g^2\mu_0^2$. The standard error is shown as bands. The smaller diagram shows the ratio of dense to dilute model scattering cross sections. As expected, the discrepancy between the dilute and the dense results increases for larger color-charge densities.}
	\label{fig:inco-g2mu02}
\end{figure}

\begin{figure}
	\centering
	\includegraphics[width=0.5\textwidth]{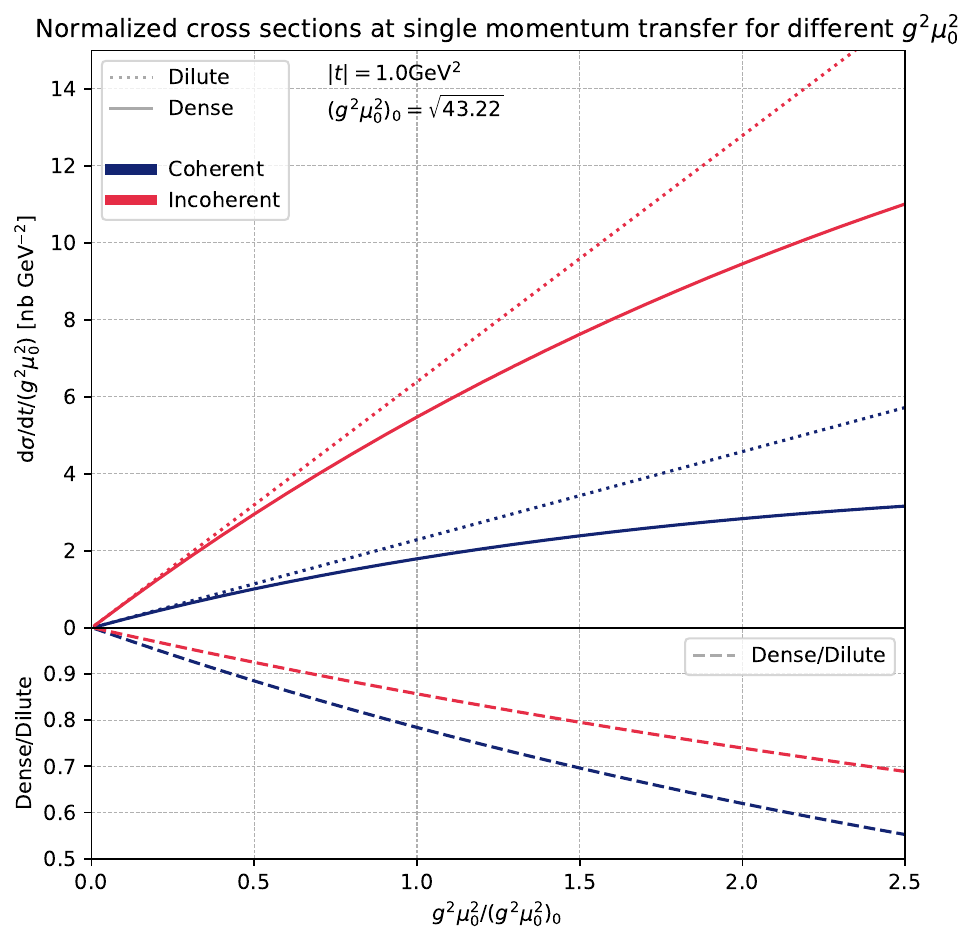}
	\caption{Coherent and incoherent cross sections at fixed energy transfer $\abs{t}=1.0\,\mathrm{GeV}^2$, for a range of different values of $g^2\mu_0^2$, compared to the value fixed for data comparison in this work, $(g^2\mu_0^2)_0 = \sqrt{43.22}$. The cross sections have been normalized by a factor $g^2\mu_0^2$ to facilitate the comparison between dense and dilute cases.}
	\label{fig:g2mu02}
\end{figure}

For the incoherent cross section in \cref{fig:inco}, we observe that the hotspot fluctuations are the leading contribution to the incoherent cross section for $\abs{t} \approx 0.5-2\,\mathrm{GeV}^2$, which is also the case of the dilute approximation in Ref.~\cite{Demirci:2022wuy}. Outside this range, they decrease swiftly, leading to the the color fluctuations dominating instead. We observe that for $|t|>2\,\mathrm{GeV}^2$, both the dense and dilute calculations approximate the analytical estimation $(1+\ln(m^2/|t|))/|t|$ of Ref.~\cite{Demirci:2022wuy}. 
While the total normalization of the coherent cross section has been fixed to successfully reproduce the H1 data, additional scaling is needed to reproduce the incoherent cross section, where we have applied a factor $K=3$ to our computation. 
It is important to note that the discrepancy between H1 and sampled results for the incoherent cross section cannot easily be corrected without affecting the coherent cross section as well. For example, if one increases the value of the dimensionless free parameters $g^2\mu_0^2$, this increases the total normalization of both the coherent and incoherent cross sections. In the case of the dense limit, the cross sections depend non-linearly on $g^2\mu_0^2$, leading to a steeper spectrum for larger values. This is demonstrated in \cref{fig:co-g2mu02,fig:inco-g2mu02,fig:inco-g2mu02-fluc,fig:g2mu02}, where we have investigated the dependence that the vector meson production cross sections have on the color charge density normalization, $g^2\mu_0^2$.
While this behavior adds to the trend observed in previous studies~\cite{Lappi:2020ufv,Demirci:2022wuy}, recent work suggests that including the relativistic corrections to the vector-meson wave-functions, as well as fluctuations of $Q_s$ for each hotspot individually, alleviates the mismatch between H1 data and phenomenological results~\cite{Mantysaari:2025idf}.

 The effective steepening of the spectrum is exemplified by the observables sensitive to geometrical fluctuations, e.g., coherent cross section and the incoherent hotspot fluctuations,  shown in \cref{fig:co-g2mu02,fig:inco-g2mu02-fluc}. On the other hand, color fluctuations in the incoherent cross sections are virtually not affected by change in the effective size, where the ratios between dense systems and the dilute (which is affected only by normalization) is quite stable along the $|t|$ direction. In \cref{fig:inco-g2mu02}, however, we see that the total incoherent cross section scales relatively cleanly with changes to $g^2\mu_0^2$. This comes as a consequence of color fluctuations dominating the incoherent spectrum everywhere outside of the maximum amplitude of the hotspot-fluctuations spectrum.

In \cref{fig:g2mu02}, we display coherent and incoherent cross-section results against different values of $g^2\mu_0^2$, partially normalized by $g^2\mu_0^2$. Due to the leading-twist approximations in the dilute limit, cross sections depend linearly on $(g^2\mu_0^2)^2$. We present the partially normalized cross section to facilitate the observation of the breaking of this scaling due to the all-twist (dense) result. As can be readily seen, dense targets display stronger dampening due to multiple scatterings, leading to a large suppression for larger color-charge normalization parameters.

\begin{figure}[t]
    \includegraphics[width=0.5\textwidth]{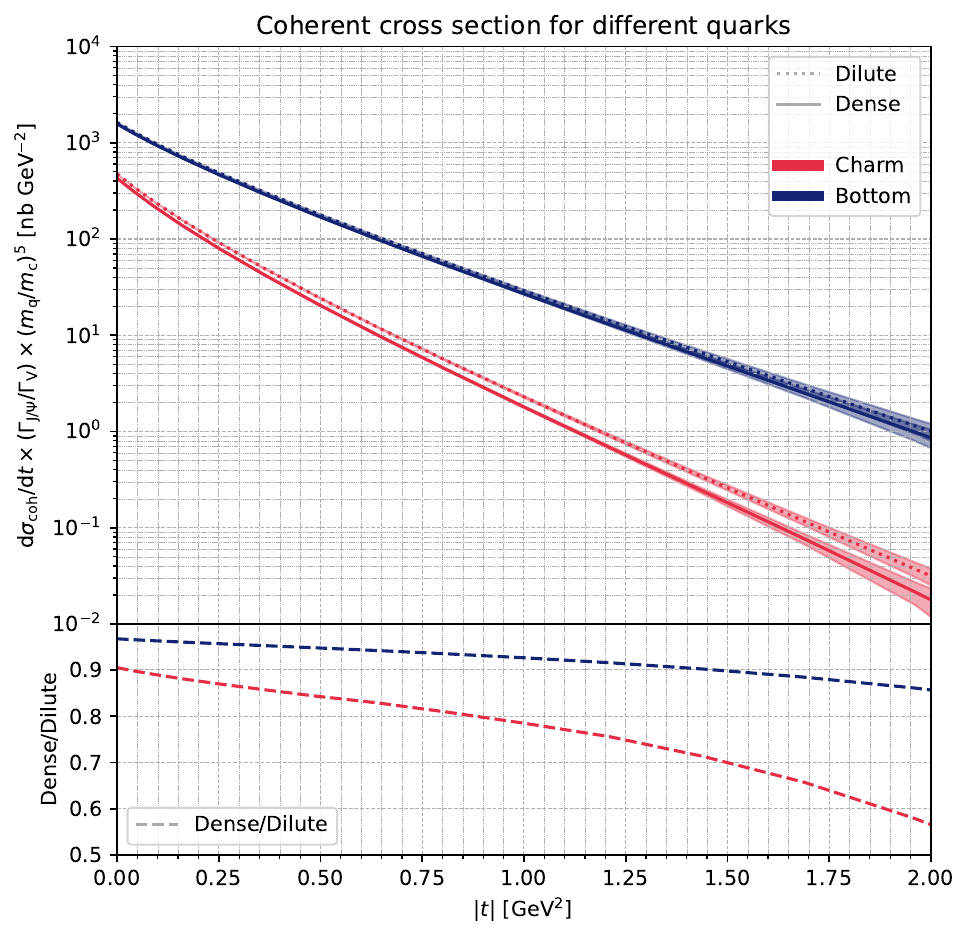}
    \caption{Coherent cross section for different quarks constituting the vector meson. Bands show the relative variation due to different event configurations.}
    \label{fig:co-quarks}
\end{figure}

\begin{figure}[t]
	\includegraphics[width=0.5\textwidth]{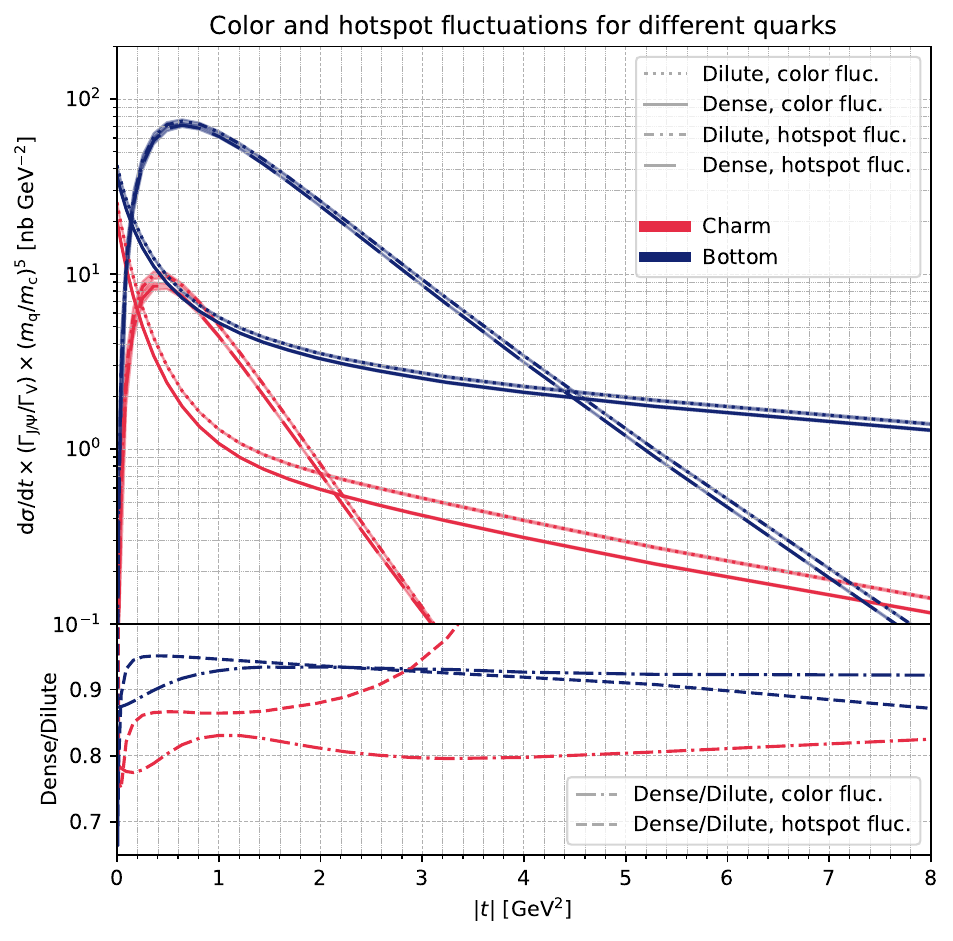}
	\caption{Color- and hotspot-fluctuation contributions making up the full incoherent cross section shown in \cref{fig:inco-quarks}, for different quarks in the vector meson. The standard error is shown as bands.}
	\label{fig:inco-quarks-fluc}
\end{figure}

\begin{figure}[t]
    \includegraphics[width=0.5\textwidth]{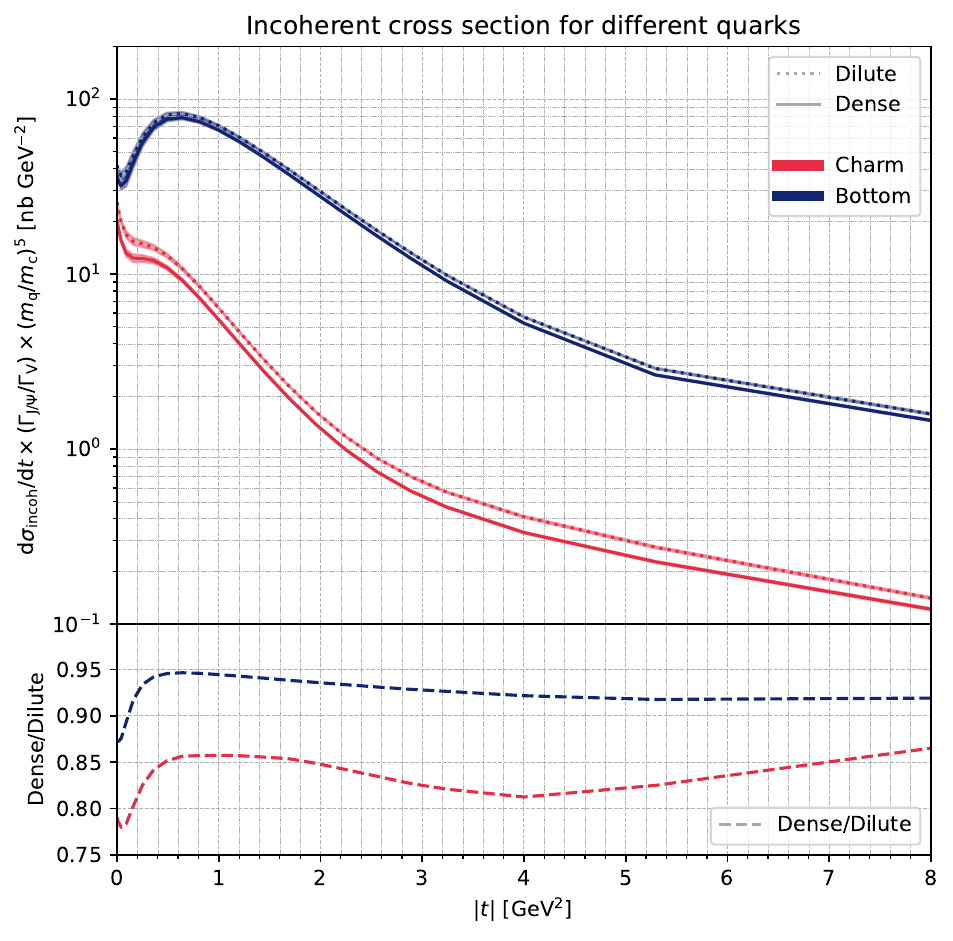}
    \caption{Incoherent cross section for different quarks constituting the vector meson. The standard error is shown as bands.}
    \label{fig:inco-quarks}
\end{figure}

The effects of the quark mass on the vector meson production coherent and incoherent cross sections are shown in \cref{fig:co-quarks,fig:inco-quarks-fluc,fig:inco-quarks}. All cross sections in these figures have been rescaled with $(\Gamma_{\mathrm{J}/\psi}/\Gamma_{V})\left(m\down{q}/m\down{c}\right)^5$ to remove the leading quark-mass dependence. The coherent cross sections, \cref{fig:co-quarks}, present a mild dependence on the quark masses, apart from the leading quark terms. The main effect is a slight change in the general decay rate of the spectrum, and hence, the effective size of the proton as seen by the $\mathrm{b}$-quark probes~\cite{Toll:2012mb}.

Considering the incoherent cross section, hotspot fluctuations are affected more than the color fluctuations by different quark masses. The hotspot spectrum sees the biggest change, as it is deeply hardened, while its amplitude relative to the color fluctuations increases. This changes the $\abs{t}$-dependence in such a way that the cross section does not steadily decrease with energy transfer but instead peaks at around $0.6\,\mathrm{GeV}^2$ for the bottom quark case. The bottom dipole forces the system into smaller dipoles, and hence large momentum transfers are enabled, hardening the spectrum. It is interesting to note that for the more massive bottom quark, the dilute and dense limits have the same slope for the momentum-transfer scales explored, which stands in contrast to the trend observed for the charm quark. On the other hand, the spectrum of color fluctuations is affected in a milder way, where the difference between targets in the dilute and dense limits is larger for less massive quarks. Additionally, just as in the case of the hotspot fluctuations, more massive quarks in the bound pair harden the spectrum by forcing the system into smaller dipoles.

\section{Conclusions} \label{sec:conclusion}

In this study we compute exclusive vector meson production cross sections in the dipole picture, using a Monte-Carlo sampling approach to account, event-by-event, for the geometrical subnuclear fluctuations of the proton, as given by the successful hotspot model. We benchmarked our numerical Monte-Carlo setup against analytical leading-twist results presented in previous work~\cite{Demirci:2022wuy}, and compared the leading- and full-twist limits to study saturation effects. For the relevant observables, we computed the diffractive coherent and incoherent cross sections of charmed and bottom vector mesons ($\mathrm{J}/\psi$ and $\Upsilon$), where we made use of the non-relativistic approximation of the wave-function overlap for the relevant vector meson. We find good agreement to coherent $\mathrm{J}/\psi$ production data from H1~\cite{H1:2003ksk,H1:2013okq}, as well as the shape of incoherent data. However, we confirm the trend observed in previous studies which indicated that an extra prefactor, $K=3$~\cite{Demirci:2022wuy}, is needed to account for the mismatch of the relative normalization of coherent and incoherent cross-sections. Since saturation effects fail to explain this mismatch, the only remaining possibility is that a more accurate description of the vector meson wave-function~\cite{Mantysaari:2025idf} beyond the strict non-relativistic limit considered in this work as well as in the leading-twist analysis of \cite{Demirci:2022wuy} is required, and we intend to pursue such an analysis in the future.

We also undertook a systematic study, exploring how the overall normalization of charge density, as well as the magnitude of the quark masses, affects the cross sections in both the dense and dilute limits. We observe a consistent trend, where larger charge densities and smaller masses are better suited to distinguish the linear and non-linear effects.

We also present a consistent parametrization of the dipole amplitude, where the color fields inside the proton are created by individual, localized charge hotspots. This is necessary to obtain a physical incoherent cross section, since simpler models with no infra-red regulation present pathological behavior for higher correlation functions (e.g. quadrupole) and large dipole sizes, especially when $\abs{\rt}>2R\down{p}$. The formulas we present here have been cured of this pathological behavior, giving a finite incoherent cross section for all geometry configurations.

Just as expected, we find that saturation effects lead to a suppression of the overall scattering cross section relative to the leading-twist limit. While being quite mild in $\mathrm{e}\mathrm{p}$ collisions, we expect, naturally, that these saturation effects are more significant in systems with higher charge densities, such as those with greater center-of-mass energies or $\mathrm{e}\mathrm{A}$ collisions. As an outlook, in future work we plan to tackle the question of how system size and nuclear structure affect the relations between the nonlinear and linear limits of vector meson production both in Ultra-peripheral collisions in RHIC and the LHC~\cite{Klein:2020nvu,Baltz:2007kq}  and the upcoming EIC~\cite{Accardi:2012qut}.

\acknowledgements
The authors would like to thank Heikki Mäntysaari and Björn Schenke for a very enlightening discussion. Additionally, OGM would like to thank Giuliano Giacalone for his interest in this work.
OGM and SS acknowledge support by the Deutsche Forschungsgemeinschaft (DFG, German Research Foundation) through the CRC-TR 211 ‘Strong-interaction matter under extreme conditions’-project number 315477589 – TRR 211. OGM and SS acknowledge also support by the German Bundesministerium für Bildung und Forschung (BMBF) through Grant No. 05P21PBCAA.
The authors acknowledge the computing time provided to them on the high-performance computers \textit{Noctua 1/2} at the NHR Center PC2. These are funded by the Federal Ministry of Education and Research and the state governments participating on the basis of the resolutions of the GWK for the national high-performance computing at universities (\href{https://www.nhr-verein.de/en/our-partners}{nhr-verein.de/en/our-partners}). 
\appendix
\section{Dipole cross section and color correlations}
\label{app:correlations}
Beginning with the definition of the dipole correlator
\begin{align}
    D (\bt,\rt) = \frac{1}{\Nc} \mathrm{tr} \left[ V(\xt) V^\dagger(\yt) \right]\,,
\end{align}
first the color average is applied and then exchanged with the trace, which makes use of the linearity of the trace, as
\begin{align}
    \avg{D (\bt,\rt)} = \frac{1}{\Nc} \mathrm{tr} \left[ \bigavg{V(\xt) V^\dagger(\yt)} \right]\,.
\end{align}
For ease of notation, the color field operator $A_\xt$ is defined such that the Wilson line operator takes the form
\begin{align}
    V (\xt) & = P_+ \exp \left\{ ig \int^\infty_{-\infty} \rmd z^+ A^-_a (z^+,\xt)t^a \right\}
    \nn \\ & = \rme^{ig A_\xt}\,.
\end{align}
The Wilson lines in the correlator are now expanded to second order, which is sufficient for a full solution with the highly Lorentz-contracted target~\cite{Gelis:2010nm,Demirci:2021kya}, yielding
\begin{align}
    \avg{D (\bt,\rt)} & \nn \\
    \approx \frac{1}{\Nc} \mathrm{tr} & \left[ \bigggavg{\left( 1 + igA_\xt - \frac{g^2}{2}A_\xt^2 \right) \left( 1 - igA_\yt - \frac{g^2}{2}A_\yt^2 \right)} \right]\,.
\end{align}
Multiplying out the terms and using that odd order color fluctuations average to $0$, we get:
\begin{align}
    \avg{D (\bt,\rt)} \approx \frac{1}{\Nc} \mathrm{tr} \left[ 1 + g^2\avg{A_\xt A_\yt} - \frac{g^2}{2}\avg{A_\xt^2} - \frac{g^2}{2}\avg{A_\yt^2} \right]
\end{align}
Since the dilute model assumes only one scattering of the dipole on the target, we can, without loss of generality, generalize that interaction to occur at a single longitudinal $z^+$. The integration then becomes trivial and we substitute the color field operator $A_\xt$ with its generator $t^a$ (with color $a$) and corresponding stochastic variable $\xi^a_\xt$, which is now independent of $z^+$:
\begin{align}
    \avg{D (\bt,\rt)} & \approx \frac{1}{\Nc} \mathrm{tr} \biggl[ 1 + g^2 t^a t^b\bigavg{\xi^a_\xt \xi^b_\yt} - \frac{g^2}{2}t^a t^b\bigavg{\xi^a_\xt \xi^b_\xt} \nn \\
    & \quad \quad - \frac{g^2}{2}t^a t^b\bigavg{\xi^a_\yt \xi^b_\yt} \biggr]
\end{align}
The stochastic variables, of course, have the same average and correlation relation as the color field operators, namely
\begin{align}
    \bigavg{\xi_\xt^a} = 0\qquad \text{and} 
\qquad    \bigavg{\xi_\xt^a \xi_\yt^b} = \frac{\lambda_{\xt\yt}}{g^2 \CF} \delta^{ab}\,.
\end{align}
Here, $\lambda$ is a short-hand for
\begin{align}
    \lambda_{\xt \yt} = g^2 \CF \bigavg{A_a^-(z^+=0, \xt) A_a^-(\ob{z}^+=0, \yt)}\,,
  \label{eq:lambda}
\end{align}
where, to solve the $z^+$ and $\ob{z}^+$ integrals, we used that the interaction occurs at a single $z^+=0$. From the Fierz identity of the Lie algebra generators $t^a$ in adjoint color space, it follows that~\cite{Nishi:2004st}
\begin{align}
    t^a_{ij} t^a_{kl} = \frac{1}{2}\delta_{il}\delta_{jk} - \frac{1}{2\Nc}\delta_{ij}\delta_{kl}\,.
\end{align}
Thus, the trace over two generators of the same color is $\mathrm{tr} \left[ t^a t^a \right] = t^a_{ij} t^a_{ji} = \frac{1}{2}(\Nc^2-1)$. Evaluating the trace and averages, we find the color averaged dipole correlator to be
\begin{align}
    \avg{D (\bt,\rt)} \approx 1 + G_{\xt \yt}\,,
\end{align}
where the dipole correlation function
\begin{align}
    G_{\xt \yt} = \lambda_{\xt \yt} - \frac{1}{2}\lambda_{\xt \xt} - \frac{1}{2}\lambda_{\yt \yt}
\end{align}
was defined. This will be calculated in Appendix~\ref{app:G}.

This leads to the dipole cross section, in the dilute limit, taking the form
\begin{align}
    \biggggavg{\frac{\rmd\sigma\up{p,dilute}\down{dip}}{\rmd^2\bt} (\bt, \rt)} = 2 \left( 1 - \avg{D(\bt, \rt)} \right) \approx -2 G_{\xt \yt}\,,
\end{align}
where, as before, $\xt = \bt + \frac{\rt}{2}$ and $\yt = \bt - \frac{\rt}{2}$.

\subsection{Dipole cross section in the dense limit}
We again start from the trace over the color average of the dipole correlator. In the dense limit, interactions between probe and target are assumed to occur all the way along the path through the target, and the integral will not become trivial. Instead, we again introduce stochastic variables $\xi$, but this time carry out the Gaussian averaging of the Wilson line correlators by representing the standard Gaussian integral over color charges as a discrete stochastic process in the evolution variable $z$. Wilson lines are initialized such that $V(z=0) = 0$, and in each evolution step, we expand the individual Wilson lines as follows:
\begin{align}
    V_\xt&(z + \delta z) \nn \\
    & = V_\xt(z) \left( 1 + ig t^a \xi^a_\xt (\delta z)^{\tfrac{1}{2}} - \frac{g^2}{2}t^a \xi^a_\xt t^b \xi^b_\xt \delta z+ ... \right)
\end{align}
Here, $\delta z$ is size of each step. Expanding both Wilson lines in this way yields
\begin{align}
    V_\xt(z&+\delta z)V^\dagger_\yt(z+\delta z) \nn \\
    & = V_\xt(z) \left( 1 + ig t^a \xi^a_\xt (\delta z)^{\tfrac{1}{2}} - \frac{g^2}{2}t^a \xi^a_\xt t^b \xi^b_\xt \delta z+ ... \right) \nn \\
    & \quad \left( 1 + ig t^c \xi^c_\yt (\delta z)^{\tfrac{1}{2}} - \frac{g^2}{2}t^c \xi^c_\yt t^d \xi^d_\yt \delta z+ ... \right)^\dagger V^\dagger_\yt(z) \\
    & = V_\xt(z) \Bigl(1 + g^2 t^a \xi_\xt^a t^c \xi_\yt^c \delta z \nn \\
    & \quad - \frac{g^2}{2} t^a \xi_\xt^a t^b \xi_\xt^b \delta z - \frac{g^2}{2} t^c \xi_\yt^c t^d \xi_\yt^d \delta z \nn \\
    & \quad + O(g^1) + O(g^3) + ...\Bigr) V^\dagger_\yt(z)\,.
\end{align}
All odd orders and orders above the second are disregarded. When now applying the color average and trace, as well as equivalently rearranging the equation and dividing by $\delta z$, we get a finite-differences differential equation:
\begin{align}
    \partial_z &\left(V_\xt(z) V^\dagger_\yt(z)\right) \nn \\
    & = \lim_{\delta z \rightarrow 0} \frac{V_\xt(z+\delta z)V^\dagger_\yt(z+\delta z) - V_\xt(z) V^\dagger_\yt(z)}{\delta z} \\
    & = V_\xt(z) \Bigl( g^2 t^a \xi_\xt^a t^c \xi_\yt^c - \frac{g^2}{2} t^a \xi_\xt^a t^b \xi_\xt^b \nn \\
    & \qquad \qquad - \frac{g^2}{2} t^c \xi_\yt^c t^d \xi_\yt^d \Bigr) V^\dagger_\yt(z)
\end{align}

We continue by calculating the $z$-derivative of our color-averaged dipole correlator, making use of the commutativity of the derivative with both the average and the trace. Using the Fierz identity for the Lie-group generators, as is shown above for the calculation of the dilute-limit correlator, and applying the average with Wick's theorem, yields the differential equation
\begin{align}
    \partial_z \avg{D(\bt, \rt)} & = \frac{1}{\Nc} \mathrm{tr} \left[ \biggavg{\partial_z \left( V_\xt(z) V^\dagger_\yt(z) \right) } \right] \\
    & = G_{\xt \yt} \avg{D(\bt, \rt)}\,,
\end{align}
which is solved by
\begin{align}
    \avg{D(\bt, \rt)} = \rme^{G_{\xt \yt}}\,.
\end{align}
Here, the initial condition of $D(\bt,\rt)(z=0) = 1$ was chosen, as $D$ is already normalized in such a way that the dipole cross section is $0$ before interaction with the target.
Thus, the dipole cross section in the dense limit is given by
\begin{align}
    \biggggavg{\frac{\rmd\sigma\up{p,dense}\down{dip}}{\rmd^2\bt} (\bt, \rt)} = 2 \left( 1 - \rme^{G_{\xt \yt}} \right)\,.
\end{align}

\subsection{Dipole-dipole cross section in the dilute limit}
Starting from the color average of the dipole-dipole cross section, we find
\begin{align}
    \biggggavg{ \frac{\rmd\sigma\up{p,dilute}\down{dip}}{\rmd^2\bt}&(\bt,\rt) \frac{\rmd\sigma\up{p,dilute}\down{dip}}{\rmd^2\ob{\bt}}(\bbt,\rbt) } \nn \\
    & = 4 \bigggavg{ \left( 1 - \frac{1}{\Nc}\mathrm{tr} \left[ V_\xt V_\yt^\dagger \right] \right) \left( 1 - \frac{1}{\Nc}\mathrm{tr} \left[ V_\xbt V_\ybt^\dagger \right] \right) }\,.
\end{align}
The traces are now expanded as they were in the single-dipole correlator calculation. Multiplying the terms and applying the trace and average analogously to the previous calculations yields
\begin{align}
    &\biggggavg{ \frac{\rmd\sigma\up{p,dilute}\down{dip}}{\rmd^2\bt}(\bt,\rt) \frac{\rmd\sigma\up{p,dilute}\down{dip}}{\rmd^2\ob{\bt}}(\ob{\bt},\ob{\rt}) } \nn \\
    & \approx 4 \left( G_{\xt \yt}G_{\xbt \ybt} + \frac{1}{2(\Nc^2-1)} \left( G_{\xt \xbt} + G_{\yt \ybt} - G_{\xt \ybt} - G_{\yt \xbt} \right)^2 \right)\,.
\end{align}

\subsection{Dipole-dipole cross section in the dense limit}
\begin{align}
    \biggggavg{ \frac{\rmd\sigma\up{p}\down{dip}}{\rmd^2\bt}&(\bt,\rt) \frac{\rmd\sigma\up{p}\down{dip}}{\rmd^2\ob{\bt}}(\ob{\bt},\ob{\rt}) }
    \nn \\ & = \biggggavg{ 2 \left( 1 - \frac{1}{\Nc}\mathrm{tr} \left[ V_\xt V_\yt^\dagger \right] \right) 2 \left( 1 - \frac{1}{\Nc}\mathrm{tr} \left[ V_\xbt V_\ybt^\dagger \right] \right) }
    \nn \\ & = 4 \bigggavg{ 1-\frac{1}{\Nc}\mathrm{tr} \left[ V_\xt V_\yt^\dagger \right] - \frac{1}{\Nc}\mathrm{tr} \left[ V_\xbt V_\ybt^\dagger \right] \nn \\
    & \quad +\frac{1}{\Nc^2}\mathrm{tr} \left[ V_\xt V_\yt^\dagger \right] \mathrm{tr} \left[ V_\xbt V_\ybt^\dagger \right]}\,.
\end{align}
Analogous to the calculation in the dilute limit, only the dipole-dipole correlator is unknown. Using the same assumptions and properties as in the calculation of the single dipole correlator in the dense model, namely the discretization of the integral as a stochastic process, we find the differential equation
\begin{align}
    \partial_z \biggavg{&\mathrm{tr}\left[ V_\xt V_\yt^\dagger \right] \mathrm{tr}\left[ V_\xbt V_\ybt^\dagger \right]} \nn \\
    & = \left( G_{\xt \yt} + G_{\xbt \ybt} - \frac{1}{\Nc^2-1}T_{\xt \yt, \xbt \ybt} \right) \biggavg{\mathrm{tr}\left[ V_\xt V_\yt^\dagger \right] \mathrm{tr}\left[ V_\xbt V_\ybt^\dagger \right]} \nn \\
    & \quad + \frac{1}{2\CF} T_{\xt \yt, \xbt \ybt} \biggavg{\mathrm{tr}\left[ V_\xt V_\ybt^\dagger V_\xbt V_\yt^\dagger \right]}
\end{align}
with the unknown quadrupole correlator
\begin{align}
    \biggavg{\mathrm{tr}\left[ V_\xt V_\ybt^\dagger V_\xbt V_\yt^\dagger \right]}\,.
\end{align}
$T$ is a transition function defined as
\begin{align}
    T_{\xt_1 \xt_2, \xt_3 \xt_4} = G_{\xt_1 \xt_4} + G_{\xt_2 \xt_3} - G_{\xt_1 \xt_3} - G_{\xt_2 \xt_4}\,.
\end{align}

Also analogous to the calculation of the dipole-dipole correlator, the $z$ derivative of the quadrupole correlator is found to be
\begin{align}
    \partial_z \biggavg{\mathrm{tr}&\left[ V_\xt V_\ybt^\dagger V_\xbt V_\yt^\dagger \right]} \nn \\
    & = \left( G_{\xt \ybt} + G_{\xbt \yt} - \frac{1}{\Nc^2-1} T_{\xt \ybt, \xbt \yt} \right) \biggavg{\mathrm{tr}\left[ V_\xt V_\ybt^\dagger V_\xbt V_\yt^\dagger \right]} \nn \\
    & \quad + \frac{1}{2\CF} T_{\xt \ybt, \xbt \yt} \biggavg{\mathrm{tr}\left[ V_\xt V_\yt^\dagger \right] \mathrm{tr}\left[ V_\xbt V_\ybt^\dagger \right]}\,.
\end{align}
The two differential equations are combined into the coupled equation
\begin{align}
    \partial_z &  \begin{pmatrix}
                    \biggavg{\mathrm{tr}\left[ V_\xt V_\yt^\dagger \right] \mathrm{tr}\left[ V_\xbt V_\ybt^\dagger \right]} \\
                    \biggavg{\mathrm{tr}\left[ V_\xt V_\ybt^\dagger V_\xbt V_\yt^\dagger \right]}
                \end{pmatrix} \nn \\
                & \quad \quad \quad \quad \quad \quad = M_{\xt \yt \xbt \ybt} \begin{pmatrix}
                                            \biggavg{\mathrm{tr}\left[ V_\xt V_\yt^\dagger \right] \mathrm{tr}\left[ V_\xbt V_\ybt^\dagger \right]} \\
                                            \biggavg{\mathrm{tr}\left[ V_\xt V_\ybt^\dagger V_\xbt V_\yt^\dagger \right]}
                                        \end{pmatrix}\,,
\end{align}
where the correlation matrix $M$ was defined as
\begin{align}
    &M_{\xt \yt \xbt \ybt} \equiv \begin{pmatrix}
    	a & b\\
    	c & d
    \end{pmatrix} \nn\\
        & ~ = \begin{pmatrix}
            G_{\xt\yt}+G_{\xbt\ybt}-\frac{1}{N\down{c}^2-1}T_{\xt\yt,\xbt\ybt} & \frac{1}{2C\down{F}}T_{\xt\yt,\xbt\ybt} \\
            \frac{1}{2C\down{F}}T_{\xt\ybt,\xbt\yt} & G_{\xt\ybt}+G_{\xbt\yt}-\frac{1}{N\down{c}^2-1}T_{\xt\ybt,\xbt\yt}
        \end{pmatrix}\,.
\end{align}
The matrix elements $a$, $b$, $c$, and $d$ were defined for clarity and will be used in the next step.

The coupled equation is solved for the dipole-dipole correlator by
\begin{align}
    \biggavg{\mathrm{tr}\left[ V_\xt V_\yt^\dagger \right] \mathrm{tr}\left[ V_\xbt V_\ybt^\dagger \right]} = \begin{pmatrix}
            1 \\
            0
        \end{pmatrix}\up{T} \rme^{M_{\xt \yt \xbt \ybt}} \begin{pmatrix}
                                                            \Nc^2 \\
                                                            \Nc
                                                        \end{pmatrix}\,.
\end{align}
The matrix exponential for a square matrix is readily calculated, such that the correlator is found to be
\begin{align}
    & \biggavg{\mathrm{tr}\left[ V_\xt V_\yt^\dagger \right] \mathrm{tr}\left[ V_\xbt V_\ybt^\dagger \right]}\nn\\
    & \qquad = \frac{\Nc^2}{2} \rme^{\frac{a+d}{2}} \left[(1+F) \rme^{\frac{1}{2}f} + (1-F) \rme^{-\frac{1}{2}f} \right]\,,
\end{align}
where $f = \sqrt{4bc+(a-d)^2}$ and $F = \frac{a-d+\frac{2b}{\Nc}}{f}$.

\noindent \textbf{Non-cancellations:} A well-used approximation is to change the full $G_{\xt\yt}$ function to an (semi-) analytical model, such as GBW or the IP-Sat models. 
As stated in the main text, we observe inconsistencies in the resulting $N$-point functions when these parametrizations are applied. In our case, this leads the quadrupole correlation to include incomplete cancellations, thus yielding divergent amplitudes due to residual non-zero value, even in the limit $|\bt| \to \infty$. In such parametrizations, $G_{\xt\yt} 
\to (1/4) \,\rt^2 Q^2(x,\rt,\bt)$, where the impact parameter dependence is then further factorized to $Q^2(x,\rt,\bt):=Q^2_0(x,\rt) \,T(\bt)$. 
This factorization, which is not present in \cref{eq:GXYR}, introduces the alleged non-cancellations. This will be confirmed by using the MV model in its non-impact parameter form~\cite{Mantysaari:2024zxq}, where the logarithm of the dipole function also factorizes, giving
\begin{equation}
G_{\bf xy} = -\frac{g^2\mu_0^2\,C_F}{4\pi} T(\mathbf{b})\frac{1}{m^2}
\left[1-m r K_1 (mr)\right]
\end{equation}
where $K_1(x)$ is a modified Bessel function of the second kind. One can obtain this infinitely large ion formula by expanding to leading order on the overlap between $\mathbf{r}$, and $\mathbf{b}$. In this sense, it acts as the first order of a gradient expansion on the colour source density of the target.

\begin{figure}
    \centering
    \includegraphics[width=1\linewidth]{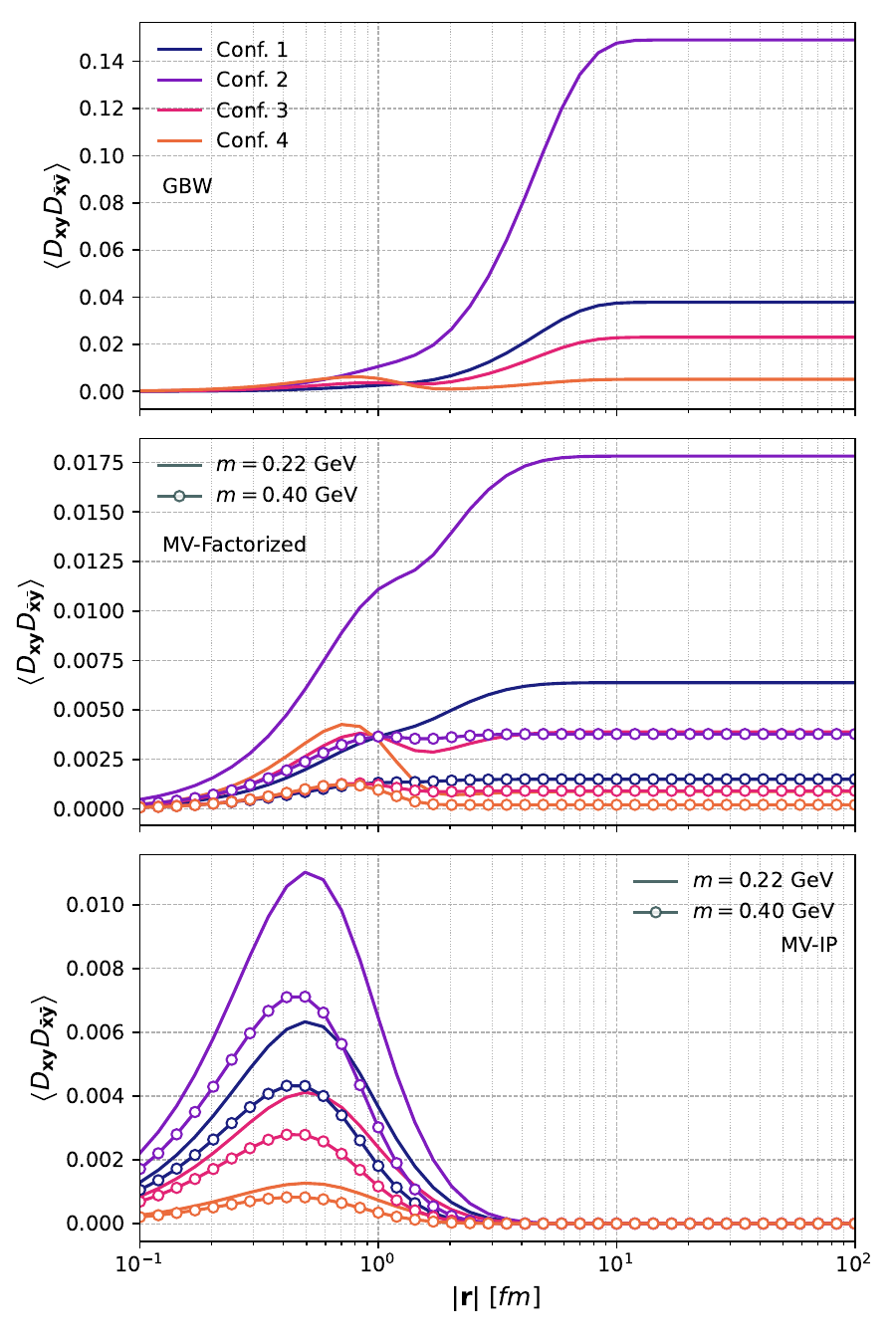}
    \caption{Dipole-Dipole correlations in terms of the first dipole size,$r$ for a set of four different outlier dipole configurations (see main text for details). }
    \label{fig:noncancellations}
\end{figure}

To make explicit the non-cancellations, we present a  simple but intuitive comparison of four different two-dipole outlier configurations, see \cref{fig:noncancellations}. In the following configurations, and for simplicity, the first dipole is placed "outside" the effective area of the gaussian profile, at $\mathbf {b} =(0,  2 r_H)$. Both dipoles are oriented in the x direction, $\mathbf{r} = (r,0)$.  \begin{enumerate}
\item Second dipole with a fixed size $\bar{r}=r_H$ located at the center of the gaussian profile, $\mathbf{\bar{b}}=0$.
\item Second dipole with a fixed size $\bar{r}=2\,r_H$ located at the center of the gaussian profile, $\mathbf{\bar{b}}=0$.
\item Second dipole with a fixed size $\bar{r}=r_H$ located opposite to the first dipole,  $\mathbf{\bar{b}}=(0,-2\,r_H)$.
\item Second dipole with a fixed size $\bar{r}=r_H$ located opposite to the first dipole, with a net imbalance. Namely,  $\mathbf{\bar{b}}=(0,-\,r_H)$.
\end{enumerate}

The reader can see that the dipole-dipole correlation stays 
In the limit of factorization of the impact parameter dependence, the incoming dipoles see the target effectively as an infinite nucleus with color density given at the impact parameter value, $\mu^2(\mathbf{b})$. This means that independent charges, the (anti-) quark, will see no effects from the profile itself. In other words, the target exhibit incosistent large distance tails.  This is the reason that ultimately leads to the non-cancellations. 
While this approximation has been quite successfully used to model a plethora of systems, including DIS \cite{Ducloue:2015gfa,Mantysaari:2022sux}, the effect of the odderon in scattering of various systems \cite{Benic:2023ybl,Benic:2025okp} and energy deposition for particle produciton in Heavy Ion Collisions \cite{Dumitru:2001ux,Drescher:2006ca,Garcia-Montero:2023gex,Garcia-Montero:2025bpn} (see \cite{Garcia-Montero:2025hys} for more references), this section shows that the hybrid dependence on the impact parameter has its limitations. 

On the other hand, the dipoles computed using the MV for the impact-parameter dependent gaussian profiles (here MV-IP) do not exhibit this problem, and all configurations vanish for large values of the first dipole size, $r$, where the first dipole "misses" the color density profile. 
It is worth noting that the variation with respect to the regulator does not change the vanishing of the correlator when $r\to \infty$, even as it has some effect on the values of the correlation, including the slope at large $r$.

\subsection{Dipole correlation function $G$} \label{app:G}
The color field $A^-_a$ was previously (\cref{eq:wilson}) defined as
\begin{align}
    A^-_a(z^+, \xt) = \int \rmd^2\zt~ G(\xt-\zt) \rho_a(z^+, \zt)\,,
\end{align}
where $G(\xt-\zt)$ is the Green's function
\begin{align}
    G(\xt-\zt) = \int \frac{\rmd^2\kt}{(2\pi)^2} \frac{\rme^{\rmi\kt(\xt-\zt)}}{\kt^2+m^2}\,.
\end{align}
In the calculation of the dipole and dipole-dipole correlators, we defined the dipole correlation function as
\begin{align}
    G_{\xt \yt} = \lambda_{\xt \yt} - \frac{1}{2}\lambda_{\xt \xt} - \frac{1}{2}\lambda_{\yt \yt}
\end{align}
and its components as
\begin{align}
    \lambda_{\xt \yt} = g^2 \CF \bigavg{A_a^-(\xt) A_a^-(\yt)}\,.
\end{align}
In the eikonal limit all interactions with the target occur on a single instant in light-cone time, and hence, in what follows, the $z^+$- and $\ob{z}^+$-dependencies of $A^-_a(z^+,\xt)$ and $A^-_a(z^+, \yt)$ can be conveniently disregarded. Substituting these with their Green's functions and charge densities $\rho(\zt)$ yields
\begin{align}
    \lambda_{\xt \yt} = g^2 \CF &\int \rmd^2\zt~ \rmd^2\zbt~ \frac{\rmd^2\kt~\rmd^2\kbt}{(2\pi)^4} \frac{\rme^{\rmi\kt(\xt-\zt)}}{\kt^2+m^2} \frac{\rme^{\rmi\kbt(\yt-\zbt)}}{\kbt^2+m^2} \bigavg{\rho_\zt \rho_\zbt}\,.
\end{align}
The color average was moved into the integrals around the color-charge densities. The complete dipole correlation function thus takes this form:
\begin{align}
    G_{\xt \yt} &= g^2 \CF\int \rmd^2\zt~ \rmd^2\zbt~\int \frac{\rmd^2\kt}{(2\pi)^2}~ \frac{\rmd^2\kbt}{(2\pi)^2}~\bigavg{\rho_\zt \rho_\zbt} \nn \\
    & \quad \times \frac{1}{(\kt^2+m^2) (\kbt^2+m^2)}\biggl[ \rme^{\rmi\kt(\zt-\xt)}\rme^{\rmi\kbt(\zbt-\yt)} \nn \\
    & \quad \quad - \frac{1}{2}\rme^{\rmi\kt(\zt-\xt)}\rme^{\rmi\kbt(\zbt-\xt)}
    - \frac{1}{2}\rme^{\rmi\kt(\zt-\yt)}\rme^{\rmi\kbt(\zbt-\yt)} \biggr]
\end{align}
As above, $\avg{\rho_\xt \rho_\yt}$ is the color-averaged correlated hotspot color-charge density, which is given by
\begin{align}
    \avg{\rho_\xt \rho_\yt} = \sum_{i=1}^{\NH} \mu^2\left( \frac{\xt+\yt}{2} - \bt_i \right) \delta^{(2)}(\xt-\yt)
\end{align}
with
\begin{align}
    \mu^2(\xt) = \frac{\mu_0^2}{2\pi r\down{h}^2} \rme^{-\frac{\xt^2}{2 r\down{h}^2}}\,.
\end{align}
Performing the $\zt$ and $\zbt$ integration, which makes use of the $\delta$ distribution after exchanging the sum over hotspots and the integrals, yields
\begin{align}
    G_{\xt \yt}
    & = \frac{g^2\mu_0^2 \CF}{(2\pi)^4} \sum_{j=1}^{\NH} \int \rmd^2\kt~ \rmd^2\kbt~ \frac{\rme^{-\frac{r\down{h}^2}{2} \left( \kt + \kbt \right)^2 }\rme^{\rmi(\kt+\kbt)\bt_j} }{(\kt^2+m^2) (\kbt^2+m^2)} \nn \\
    & \times\left[ \rme^{-\rmi\kt\xt}\rme^{-\rmi\kbt\yt} -\frac{1}{2}\rme^{-\rmi\kt\xt}\rme^{-\rmi\kbt\xt} -\frac{1}{2}\rme^{-\rmi\kt\yt}\rme^{-\rmi\kbt\yt} \right]\,.
\end{align}
The phase $\rme^{\rmi(\kt+\kbt)\bt_j}$ is part of the result of the Fourier transformation. Factorizing this phase with the other exponentials results in
\begin{align}
    G_{\xt \yt} &= 
     \frac{g^2\mu_0^2 \CF}{(2\pi)^4} \sum_{j=1}^{\NH} \int \rmd^2\kt~ \rmd^2\kbt~ \frac{\exp\left\{-\frac{r\down{h}^2}{2} \left( \kt + \kbt \right)^2 \right\}}{(\kt^2+m^2) (\kbt^2+m^2)} \nn \\
     \times&\left[ \rme^{-\rmi(\kt\xt_j+\kbt\yt_j)} -\frac{1}{2}\rme^{-\rmi(\kt+\kbt)\xt_j} -\frac{1}{2} \rme^{-\rmi(\kt+\kbt)\yt_j}\right]\,,
\end{align}
where we have defined $\xt_i = \xt-\bt_i$ and $\yt_i = \yt-\bt_i$.
Thus, with multiple hotspots, the position of each hotspot is simply subtracted from the quark positions $\xt$ and $\yt$, which are integrated over in the scattering amplitude. Using Feynman's integration trick, namely
\begin{equation}
\frac{1}{(\kt^2+m^2)} = \int_0^\infty \rmd u~ \rme^{-u (\kt^2+m^2)}\,,
\end{equation}
the remaining integral is a complex Gaussian integration, which can performed to give
\begin{align}
    G_{\xt \yt}& = \frac{g^2\mu_0^2\CF}{(4\pi)^2}\sum_{j=1}^{\NH} \int_0^\infty \rmd u~ \int_0^\infty \rmd v~ \frac{\rme^{-m^2(u+v)}}{uv+\frac{r\down{h}^2}{2}(u+v)} \nn \\
    & \quad \times \Biggl[ \exp \left\{ -\frac{1}{4}\frac{u\yt_j^2+v\xt_j^2+\frac{r\down{h}^2}{2}(\xt_j-\yt_j)^2}{uv+\frac{r\down{h}^2}{2}(u+v)} \right\} \nn \\
    & \quad \quad -\frac{1}{2}\exp \left\{ -\frac{1}{4} \frac{(u+v)\,\xt_j^2}{uv+\frac{r\down{h}^2}{2}(u+v)} \right\} \nn \\
    & \quad \quad -\frac{1}{2}\exp \left\{ -\frac{1}{4} \frac{(u+v)\,\yt_j^2}{uv+\frac{r\down{h}^2}{2}(u+v)} \right\} \Biggr]\,.
\end{align}

The reader can note that $G_{\xt\yt}$ only depends on the magnitudes $\abs{\xt_i}$ and $\abs{\yt_i}$, and the angle between $\xt_i$ and $\yt_i$, $\phi_{\xt_i,\yt_i}$, since
\begin{equation}
\begin{split}
    (\xt-\yt)^2 &= (\xt_i-\yt_i)^2 \\&= \xt_i^2 + \yt_i^2 - 2\sqrt{\xt_i^2 \yt_i^2} \cos \phi_{\xt_i,\yt_i}\,.
\end{split}
\end{equation}

Because the numeric scattering amplitude integrals already have four and eight dimensions respectively, this two-dimensional integration cannot realistically be included in every integration point. An interpolation method was chosen to replace the two integrals giving $G$. This tricubic interpolation uses the magnitudes of the vectors and the angle between them as inputs to interpolate over precalculated results, achieving a significant speedup for each calculation of $G$, compared to numeric integration methods, with only small losses in accuracy (relative error $<10^{-4}$).
\begin{figure}[t]
    \centering
    \includegraphics[width=1.0\linewidth]{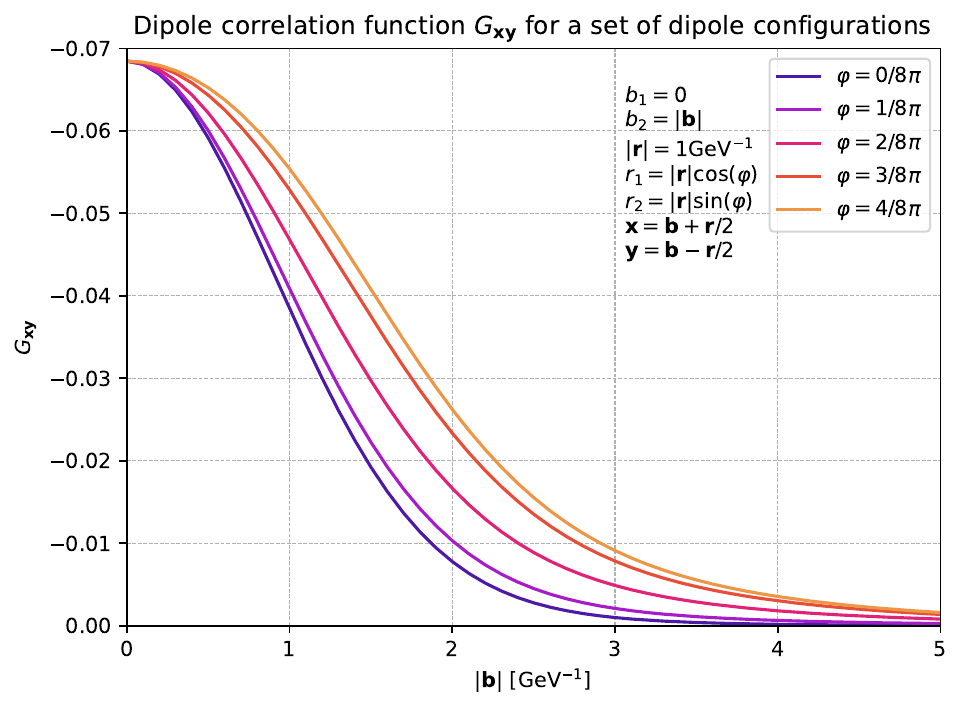}
    \caption{Dipole correlation function for different dipoles. The size of the dipole, $r=\abs{\rt}$, was arbitrarily chosen to be $1\,\mathrm{GeV}^{-1}$.}
    \label{fig:g}
\end{figure}
\cref{fig:g} shows the correlation function for one hotspot of size $r\down{h}^2=0.7\,\mathrm{GeV}^2$ at the origin and a set of different rotations of the dipole that scatters off it. More alignment of the dipole along the gradient of the hotspot's charge density results in higher dipole correlation

The largest values are found when the dipole quarks pass on opposite sides of the hotspot, as is shown at $\abs{\bt} \approx 0$.

\section{Numerics} \label{app:numerics}

\subsection{Numerical integration}
Numeric integrators~\cite{cubature} were used to solve the four- and eight-dimensional integrals required for the amplitudes in the coherent and incoherent cross sections.

All proper numeric calculations before analysis were performed on high-performance computers at the Paderborn Center for Parallel Computing (PC2)~\cite{PC2}.

\subsection{Hotspot sampling}
Generating a hotspot configuration consists of first sampling $\NH$ independent positions in transverse 2d space according to the spatial density distribution of the proton (see \cref{eq:thickness}). Then, the spatial average of these positions is calculated and subtracted from each position, which aligns the center of mass with the desired center of the proton. This process causes the actual size of the spatial distribution of hotspots to be slightly smaller than the size of the distribution on which they were sampled initially. Previous work~\cite{Demirci:2022wuy} found that, with Gaussian hotspot and proton profiles, as are in use here as well, the resulting distribution is still Gaussian, and calculated the new size parameter, the "coherent radius", to be~\cite{Demirci:2022wuy}
\begin{align}
  R\down{C}^2 = r\down{h}^2 + \frac{\NH-1}{\NH}R^2\,.
\end{align}
The Monte Carlo-sampling approach outlined in this work does not need the analytical solution for the resulting distribution.

\subsection{Energy transfer orientation average} \label{app:phi-avg}
Averaging over the polar angle $\phi_\bfDelta$ of $\bfDelta$ as
\begin{align}
    \frac{\rmd\sigma}{\rmd t} = \int \frac{\rmd\phi_\bfDelta}{2\pi}\frac{\rmd\sigma}{\rmd(-\bfDelta^2)}
\end{align}
yields the desired, direction independent cross section.
For the the color fluctuations this can be done semi-analytically:
\begin{widetext}
\begin{equation}
\begin{split}
\int \frac{\rmd\phi_\bfDelta}{2\pi} \left( \bigavg{\biabs{A(e)}^2}\down{c}-\bigabs{\avg{A(e)}\down{c}}^2 \right) 
    & = \int \frac{\rmd\phi_\bfDelta}{2\pi} \frac{1}{16\pi^2} \int \rmd^2\rt~ \rmd^2\rbt \int \rmd^2\bt~ \rmd^2\bbt~ \left( \Psi_\gamma^* \Psi\down{V} \right) (Q^2,\rt)~ \left( \Psi_\gamma^* \Psi\down{V} \right) (Q^2,\rbt) \\
    & \quad \times \rme^{ -\rmi (\bt-\bbt) \bfDelta } \Biggl( \biggggavg{\frac{\rmd\sigma\up{p}\down{dip}}{\rmd^2\bt}(\bt,\rt) \frac{\rmd\sigma\up{p}\down{dip}}{\rmd^2\bbt}(\bbt,\rbt)} - \biggggavg{\frac{\rmd\sigma\up{p}\down{dip}}{\rmd^2\bt}(\bt,\rt)} \biggggavg{\frac{\rmd\sigma\up{p}\down{dip}}{\rmd^2\bbt}(\bbt,\rbt)} \Biggr)
\end{split}
\end{equation}
Here, the first term inside the parentheses of the reduced dipole cross section is the color-averaged dipole-dipole correlator as shown above. The second term is the trivial product of two single color-averaged dipole correlators arising from the square of the amplitude before the hotspot averaging. Performing the $\phi_\bfDelta$ integral yields the following result:
\begin{equation}
\begin{split}
\int \frac{\rmd\phi_\bfDelta}{2\pi} \left( \bigavg{\biabs{A(e)}^2}\down{c}-\bigabs{\avg{A(e)}\down{c}}^2 \right) & = \frac{1}{16\pi^2} \int \rmd^2\rt~ \rmd^2\rbt ~ \int \rmd^2\bt~ \rmd^2\bbt~ \,J_0\left( \biabs{\bt-\bbt}\biabs{\bfDelta} \right) \left( \Psi_\gamma^* \Psi\down{V} \right) (Q^2,\rt)~ \left( \Psi_\gamma^* \Psi\down{V} \right) (Q^2,\rbt) \\
    &  \qquad \qquad \times\Biggl( \biggggavg{\frac{\rmd\sigma\up{p}\down{dip}}{\rmd^2\bt}(\bt,\rt) \frac{\rmd\sigma\up{p}\down{dip}}{\rmd^2\bbt}(\bbt,\rbt)} - \biggggavg{\frac{\rmd\sigma\up{p}\down{dip}}{\rmd^2\bt}(\bt,\rt)} \biggggavg{\frac{\rmd\sigma\up{p}\down{dip}}{\rmd^2\bbt}(\bbt,\rbt)} \Biggr)
\end{split}
\end{equation}
$J_0$ denotes the Bessel function of the first kind. The scattering cross section now only depends on the magnitude of $\bfDelta$.
\end{widetext}

\bibliography{References}

\end{document}